\documentclass[]{aa}
\usepackage{natbib}
\usepackage{graphicx}
\usepackage{epsf}
\begin{document}

\title{Classification and redshift estimation by principal component analysis}
\author{R\'emi A. Cabanac \inst{1,2}
\thanks{Fellow of Fonds FCAR, Qu\'ebec}
\and Val\'erie de Lapparent \inst{1}
\and Paul Hickson \inst{3}
}

\offprints{R. A. Cabanac}

\institute{
Institut d'Astrophysique de Paris, CNRS / Univ. Pierre et Marie Curie, 98 bis Bd Arago, 75014 Paris, France
\and European Southern Observatory, Vitacura, Alonso de Cordova, 3107, casilla 19001, Santiago, Chile 
\and Dept. of Physics \& Astronomy, Univ. British Columbia, 2219 Main Mall, Vancouver, BC V6T1Z4, Canada 
}
\date{Received 3 April 2001 / Accepted 22 April 2002}

\abstract{

We show that the first 10 eigencomponents of the
Karhunen-Lo\`eve expansion or Principal Component Analysis (PCA) provide
a robust classification scheme for the identification of stars, galaxies and 
quasi-stellar objects from multi-band photometry. To quantify the efficiency 
of the method, realistic simulations are performed which match the planned 
Large Zenith Telescope survey. This survey is expected to provide spectral 
energy distributions with a resolution $R\simeq40$ for $\sim10^6$ galaxies 
to $R\le23$ ($z\sim 1$), $\sim 10^4$ QSOs, and $\sim 10^5$ stars. 

We calculate that for a median signal-to-noise ratio of 6, 98\% of stars, 
100\% of galaxies and 93\% of QSOs are correctly classified. These values 
increase to 100\% of stars, 100\% of galaxies and 100\% of QSOs at a 
median signal-to-noise ratio of 10. The 10-component PCA also allows 
measurement of redshifts with an accuracy of $\sigma_\mathrm{Res.}\la0.05$ for galaxies with 
$z\la0.7$, and to $\sigma_\mathrm{Res.}\la0.2$ for QSOs with $z\ga2$, at a 
median signal-to-noise ratio of 6. At a median signal-to-noise ratio 20, 
$\sigma_\mathrm{Res.}\la0.02$ for galaxies with $z\la1$ and for QSOs with $z\ga2.5$
(note that for a median $S/N$ ratio of 20, the bluest/reddest objects will 
have a signal-to-noise ratio of $\la 2$ in their reddest/bluest filters).
This redshift accuracy is inherent to the $R\simeq40$ resolution 
provided by the set of medium-band filters used by the Large Zenith 
Telescope survey. It provides an accuracy improvement of nearly an order of 
magnitude over the photometric redshifts obtained from broad-band $BVRI$ 
photometry.

\keywords{galaxies: distances and redshifts -- galaxies: fundamental parameters -- 
methods: statistical -- quasars: general -- stars: fundamental parameters -- 
techniques: photometric}

}

\authorrunning{Cabanac, de Lapparent, Hickson}
\titlerunning{Classification and redshift estimation by PCA}
\maketitle

\section{Introduction} \label{intro}

The galaxy luminosity function (hereafter LF), 
defined as the number density of galaxies per unit
interval of luminosity is a fundamental statistical tool
required to model the formation, evolution and clustering of the galaxies.
At $z\la1$, it is well established that the LF depends
on the galaxy morphological type \citep{bingelli88,marzke98,loveday99},
and that it evolves with redshift \citep{lilly96,lin99,bromley98,lapparent02a}. 
Measurement of the LF thus requires large galaxy samples which can be separated
into several morphological, spectral or color classes and in redshift intervals.
Among the next generation redshift surveys, only 3 will be able to probe
the galaxy LF to $z\sim1$. The DEEP redshift survey using
Keck telescopes \citep{davis98} and the VIRMOS redshift survey
using the VLT \citep{lefevre98} are optimized for clustering analyses, but they
will also provide measurements of the galaxy LFs, even if the 2 surveys
shall suffer from various aperture effects and calibration difficulties.
The third survey is the Large Zenith Telescope (hereafter LZT) survey,
which is optimized for the measurements of the galaxy LFs to $z\sim1$.

An essential step in the measurement of the LF for a systematic survey 
is the classification of objects
as stars, galaxies, QSOs, etc. For the galaxies and QSOs, an estimate of
the redshift for each object is also desired.
This paper proposes and tests a classification approach based on the Principal
Component Analysis (PCA), also known as Karhunen-Lo\`eve expansion --
the underlying principles of the PCA were independently derived by
\citet{karhunen47} and \citet{loeve48}.
The PCA is a non-parametric approach which has been successfully used
for a variety of astronomical applications including stellar classification from 
photometric data \citep{deeming64,scarfe66,whitney83,whitney83b} and from spectra
\citep{storri94,ibata97,bailer98,singh98}, galaxy classification from
photometric data \citep{watanabe85} and from galaxy spectra 
\citep{connolly95a,connolly95b,galaz98,connolly99,ronen99}, and for galaxy redshift
measurements \citep{glazebrook98}; other fields of application are
solar flare observations \citep{teuber79}, asteroid spectra \citep{britt92}, 
inter-stellar medium emission lines \citep{heyer97}, gamma ray bursts 
\citep{bagoly98}, and active galaxies  \citep{mittaz90,dultzin96,turler98}. 
 
All previous spectral classification attempts using the PCA
employed either multicolor photometry (usually fewer than 10 color bins,
e.g. $UBVRIJHK$) or medium to high-resolution spectroscopy (resolution $R > 500$).
The PCA has not been tested on spectral energy distributions (SED)
with $R \sim 40$  because no such data existed until the UBC-NASA
Multi-narrowband survey \citep{hickson98a,cabanac98}. 

In this paper we use simulations based on the LZT survey parameters to
evaluate the PCA method. Section \ref{simul} describes our simulations
of mock LZT catalogues, Sect{.} \ref{method} describes the approach
used to classify the objects using the PCA, Sect{.} \ref{result}
shows the efficiency of the classification and Sect{.} \ref{redshift}
discusses the redshift accuracies which can be obtained directly from
the PCA or from a composite method similar to that described by \citet{glazebrook98}.

\section{Mock catalogues}\label{simul}

In order to be able to simulate the classification efficiency of the PCA,
we create mock catalogues which match
as closely as possible the observations of the LZT. A \emph{calibration}
of the PCA on real data is still necessary to extract sensible
physical information. The problem is addressed in another paper in preparation
\citep{cabanac02} using data from the NASA Orbital Debris Observatory (NODO;
\citealt{hickson98a}).

\subsection{The Large Zenith Telescope}\label{lzt}

\begin{table}
\begin{center}
\caption{Characteristics of the LZT}
\label{lztchar}
\begin{tabular}{rl}
\hline
\hline
Longitude&$122^{\circ}34\arcmin22.4\arcsec$\\
Latitude&$49^{\circ}17\arcmin15.5\arcsec$\\
Altitude&403 m\\
Median seeing&$0.9\arcsec$\\
Telescope diameter&6 m\\
Focal length& 10 m\\
Diameter of corrected field&$24\arcmin$\\
Detector&Thinned 2048 x 2048 CCD\\
Image scale&$0.495\arcsec$ / pixel\\
Surveyed area&$17\arcmin\times280^{\circ}\sim80~$deg$^2$\\
Integration time&64.76~sec.\\
limiting R magnitude&25.4\\
\hline
\end{tabular}
\end{center}
\end{table}

The Large Zenith Telescope \citep{hickson98b} is a 6-m liquid-mirror telescope
under construction near Vancouver, Canada.  With first-light expected in 2002, it 
will conduct drift-scan surveys of a strip of sky centered at $49^\circ 17\arcmin$
declination.
The main characteristics of the LZT are given in Table \ref{lztchar}. It is a 
zenith-pointing telescope equipped with a $2k\times2k$ thinned CCD 
($0.5\arcsec$/pix) at the prime focus, which scans a $17\arcmin\times110^\circ$ 
strip of sky at sidereal rate.
Forty medium-band filters are employed which have logarithmically-separated 
central wavelengths from 4000 \AA~ to 1 $\mu$  and one broad-band U filter.
The survey is expected to yield calibrated spectral energy
distributions (SEDs) for $\sim10^6$ galaxies to $R\le23$ ($z\sim 1$),
$\sim 10^4$ QSOs and $\sim 10^5$ stars.

\begin{figure}
\epsfysize=7cm
\centerline {\epsfbox[0 0 550 550]{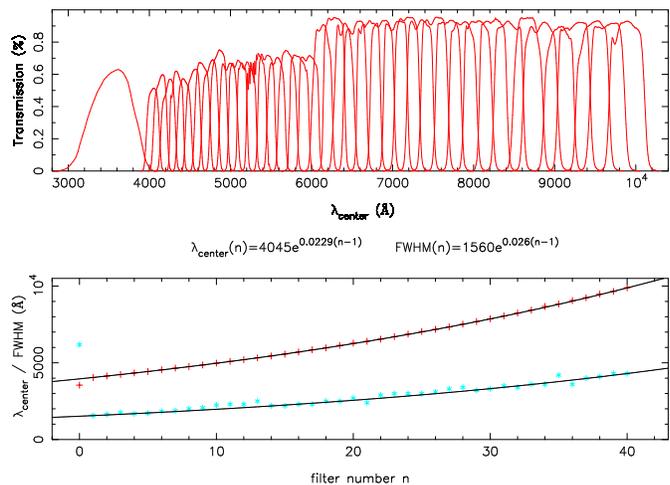}}
\caption{The transmission curves of the medium-band filters + U band used by
the LZT (top graph) and the central wavelengths $\lambda_\mathrm{center}$ (crosses) and 
$FWHM$ of the filters multiplied by 10 (bottom graph).}
\label{transcurv}
\end{figure}

\subsection{Stars}\label{stars}

\begin{table}
\begin{center}
\caption{Star counts and $B-V$ colors predicted by the Bahcall-Soneira model
of the Galaxy \protect\citep{bahcall86} for a limiting magnitude $R\le23$.
The color noise is set to 0.1 mag. The spheroid giant branch is that of M13.}
\label{tablestar}
\begin{tabular}{crrr}
\hline
\hline
& Total & Disk  & Spheroid \\
\hline
 Star counts &  3371   &   1445 &  1926 \\
 Mean $(B-V)$ &  1.07     &  1.46  &  0.76 \\
 Star  fraction &     &     43\% &  57\% \\
 Giant fraction &  12\%  & 0.4\% & 21\% \\
\hline
$(B-V)$ bin&&&\\
\hline
   0.1-0.2  &     3.3    &    0.2 &    3.1 \\
   0.2-0.3  &    31.6    &    0.7 &   30.8 \\
   0.3-0.4  &   135.7    &    2.6 &  133.0 \\
   0.4-0.5  &   277.6    &    8.6 &  268.9 \\
   0.5-0.6  &   314.3    &   21.5 &  292.7 \\
   0.6-0.7  &   264.8    &   35.0 &  229.8 \\
   0.7-0.8  &   220.8    &   37.0 &  183.7 \\
   0.8-0.9  &   183.7    &   32.4 &  151.1 \\
   0.9-1.0  &   154.7    &   29.5 &  125.1 \\
   1.0-1.1  &   146.4    &   28.8 &  117.5 \\
   1.1-1.2  &   153.8    &   32.6 &  121.2 \\
   1.2-1.3  &   166.8    &   48.4 &  118.3 \\
   1.3-1.4  &   192.0    &  101.3 &   90.6 \\
   1.4-1.5  &   265.3    &  220.9 &   44.3 \\
   1.5-1.6  &   332.2    &  319.9 &   12.2 \\
   1.6-1.7  &   270.7    &  268.5 &    2.1 \\
   1.7-1.8  &   150.7    &  150.3 &    0.3 \\
   1.8-1.9  &    69.0    &   68.9 &    0.0 \\
   1.9-2.0  &    26.8    &   26.8 &    0.0 \\
   2.0-2.1  &     8.4    &    8.2 &    0.0 \\
   2.1-2.2  &     1.7    &    1.7 &    0.0 \\
\hline
\end{tabular}
\end{center}
\end{table}

\begin{figure*}
\begin{center}
\epsfysize=15cm
\centerline {\epsfbox[51 50 583 647]{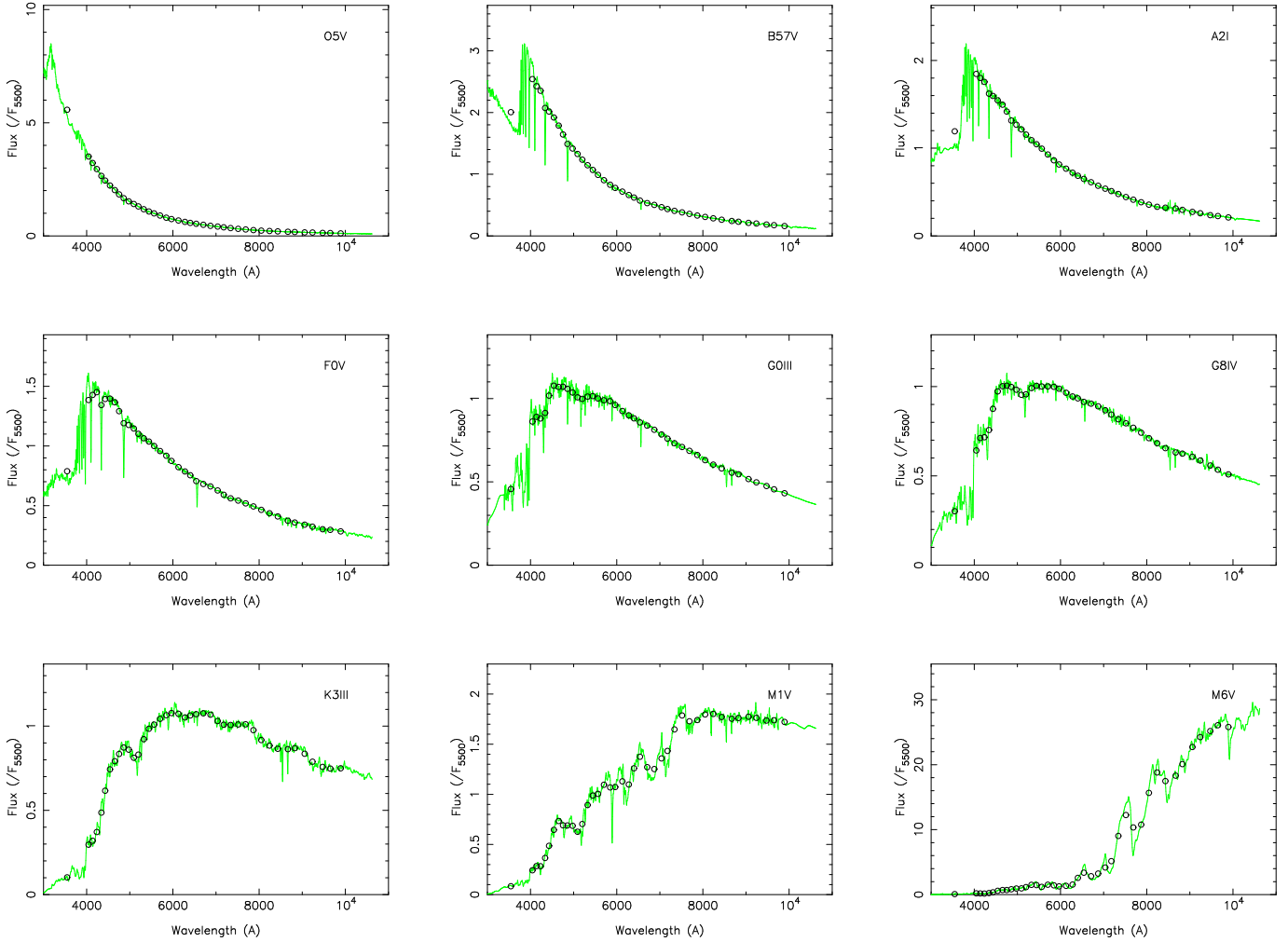}}
\caption{9 templates of stellar spectra from the Pickles library
(1998), containing 131 spectra. The high-resolution
spectra ($R\simeq 2000$) are shown (grey line) along with the LZT-filtered 
flux ($\circ$).
Spectral types are indicated.}
\label{paperspecu}
\end{center}
\end{figure*}

A phenomenological model producing the observed relative amounts of blue and red
stars is sufficient for the mock catalog. We use the galaxy model of Bahcall \& 
Soneira \citep{bahcall86}  (http://www.sns.ias.edu/$\sim$jnb/Html/galaxy.html
html) to derive star counts and colors. Table \ref{tablestar} lists the star
counts and colors predicted for a limiting magnitude of $R\le23$. The template
spectra are collected from the star catalog of \citet{pickles98},
available at CDS (http://cdsweb.u-strasbg.fr/). This catalog contains
131 stellar spectra in the range 1150-10\,620 \AA, spanning all star
temperatures (O to M), types (I to V), and metallicities (normal, rich, or
poor). The template spectra are filtered using the transmission curves of Fig{.}
\ref{transcurv}. Figure \ref{paperspecu} shows examples of template stellar
spectra.

The simulation proceeds as follows. For each color bin 
in Table \ref{tablestar}, we randomly draw the
required number of templates (number of stars given in Table \ref{tablestar})
from Pickles sub-sample of giant (or dwarf) templates. For instance,
bin $0.5<B-V<0.6$ has 22 stars in the disk component (0.4\%, that is zero 
stars are giant), and 293 stars in the spheroid component (21\% are giant, 
that is 62 stars). 
Before stacking each drawn template to the final catalog, we add noise weighted 
by the expected LZT instrumental efficiency (see Sect{.} \ref{noise}). 
The range of median signal-to-noise ratio in the continuum of the spectra which are 
generated is $6-100$.

The final mock catalog contains $\sim$ 3370 stellar templates, and faithfully 
reproduces the giant$/$dwarf fraction and the $B-V$ color distribution of Table
\ref{tablestar}.

\subsection{Galaxies}\label{galaxies}

\subsubsection{Luminosity functions}

Realistic simulations of galaxy morphology and redshift distributions
are crucial but difficult because they 
require a prior knowledge of the morphological (or spectral)
type-dependent LFs.  The ``intrinsic'' LFs based on morphological type
are only measured locally at $z\la0.03$ 
\citep{bingelli88,marzke94,jerjen97,marzke98,marinoni99}. 
Recent surveys to $z\sim0.5-1$, in which galaxies are separated in 2 or 3 classes 
using colors or spectral measures indicate that the intrinsic LFs evolve with
redshift \citep{lilly95,heyl97,lin99,lapparent02a}, but the relation
with the nearby LFs based on morphological type is not straightforward. 

To simulate an approximate LZT galaxy catalog, we must choose
a model for the intrinsic LFs in the redshift range of the LZT survey
($0\la z\la1$). Because the LZT survey will be most sensitive in the red filters, 
Table \ref{gallf} lists the galaxy LFs $\phi(M_\mathrm{R})$ of the
existing red-band selected surveys which extend at or beyond $z\sim0.5$.
The deep blue-band selected survey by \citep{heyl97} is therefore not considered.
The LFs in Table \ref{gallf} are defined by
the Schechter parameterization \citep{schechter76},
\begin{eqnarray}
\phi(M)dM~&=&~0.4~\ln{10}~\phi^*~10^{-0.4(M^*-M)(\alpha+1)}dM
\nonumber \\
&~&~~~~~~~~~~\times \left[-10^{0.4(M^*-M)}\right], \label{eqlf} \end{eqnarray}
where $M$ is the absolute magnitude, and
$\phi^*$, $M^*$, and $\alpha$ are the Schechter parameters.
All parameters in Table \ref{gallf} correspond to the
case of a flat Universe with $\Lambda=0$ and $H_0 = 100 h$ km s$^{-1}$ 
Mpc$^{-1}$.

The LFs in Table \ref{gallf} are binned into early and late-type galaxies, and
the measured evolution in the Schechter parameters is indicated for the samples
in which it is detected.
For the CNOC2 survey \citep{lin99}, the first class is their denoted 
``Early+Intermediate'' class, and the second class is their denoted ``Late'' 
class; these classes are obtained by least-square fit to the SEDs by 
\citet{coleman80}.
For the ESS \citep{lapparent02b}, the spectral classification is obtained
by a PCA calibrated on the Kennicutt sample of nearby
galaxy spectra \citep{kennicutt92}. 
\citet{lilly95} divide the
CFRS sample into a population redder than Sbc, defined as having rest-frame
$[U-V]_\mathrm{AB}=1.38$ ($[U-V]_\mathrm{AB}\simeq[V-I]_\mathrm{AB}$ at $z\sim0.5$, $V_\mathrm{AB}=V$, and
$I_\mathrm{AB}=I+0.48$), and a population bluer than Sbc, defined as the remaining 
galaxies. In Table \ref{gallf}, the Schechter LF parameter $M^*$ is listed in the $R_\mathrm{c}$ 
Cousins band, used by the CNOC2 \citep{lin99} and the ESS
\citep{lapparent02a,lapparent02b}. For the CFRS \citep{lilly95}, we use the $M^*$
values derived by the authors in the B$_\mathrm{AB}$ band for the galaxies with
$0.2\le z\le0.5$, and we convert them into the $R_\mathrm{c}$ band
using $B_\mathrm{AB}=B-0.17$ \citep{lilly95}, and assuming
$B-R_\mathrm{c}=1.35$ for galaxies redder than Sbc, and
$B-R_\mathrm{c}=0.85$ for galaxies bluer than Sbc \citep{fukugita95}.
Note that we have also converted the CFRS $M^*$ and $\phi^*$ from 
\citet{lilly95} from $H_0 = 50$ km s$^{-1}$ Mpc$^{-1}$ used by the authors,
to $H_0 = 100 h$ km s$^{-1}$ Mpc$^{-1}$ used here.

For comparison with a red-band selected survey at low 
redshift ($0\le z\le0.2$), we also quote in Table \ref{gallf}  
the LCRS survey \citep{bromley98}. The $r_\mathrm{Gunn}$ magnitudes
are converted into $R_\mathrm{c}$ using $r_\mathrm{Gunn}-R_\mathrm{c}=0.36$ \citep{fukugita95}.
For this survey, the spectral classification is obtained by a PCA, but it
is not calibrated on spectra of known morphology.  The grouping of
galaxies into early and late-types must thus be done arbitrarily.  In
Table \ref{gallf}, we show the LFs obtained by grouping galaxies in
clan $1+2+3+4$ into the early-types, and those in clans $5+6$ as the
late-types.  The corresponding listed Schechter parameters $M^*_R$ and
$\alpha$ for the early and late-type galaxies are obtained by
averaging the Schechter parameters over the considered clans. We
calculate the corresponding amplitudes $\phi^*$ for the 2 average LFs
by adjusting their integral in the absolute
magnitude interval $-23\le M_R\le-16.5$ to the sum of the observed numbers of
galaxies in the considered classes; a total survey area of 462 deg$^2$ is used
\citep{shectman96}. The systematic bias against low
surface-brightness galaxies which is present in the LCRS spectroscopic survey,
tends to exclude late-type galaxies. This 
explains the relatively low $\phi^*(z\simeq0)$ for the late-type galaxies in the
LCRS as compared to the other surveys (by a factor of 3 to 5).
This effect is might also be
present in the LCRS early-type class (defined as clans $1+2+3+4$), and could
explain the $\sim 50$\% lower $\phi^*$ compared to that for 
the deeper surveys (ESS and CFRS).

\begin{table*}
\begin{center}
\caption{Parameters of the Schechter luminosity functions measured 
by various red-band galaxy surveys with $0<z<0.5$ (see text and Eq{.} \protect\ref{eqlf}).}
\label{gallf}
\begin{tabular}{llrcccl}
\hline
\hline
Survey name    &Galaxy Type&$N_\mathrm{gal}$&$M^*_{R_\mathrm{c}}-5\log h$&$\alpha$&$\phi^*$$^\mathrm{a}$&Evolution\\
and limits     &           &         &               &        &            &         \\
\hline
CNOC2$^\mathrm{b}$  &&&&&&\\
$0.1\la z\la 0.6$  &Earlier than Sbc&1128 &$-20.61\,\pm\,0.11$ &$-0.44\,\pm\,0.10$ &$0.023$&{\scriptsize{$M^*_R(z) \simeq M^*_R(z=0.3)^\mathrm{c}$}}\\
&&&&&&~~~~~~~~~~~~{\scriptsize{$-0.7(z-0.3)$}}\\
$R_\mathrm{c}\le21.5$ &Later than Sbc  &1012 &$-20.11\,\pm\,0.18$ &$-1.34\,\pm\,0.12$ &$0.006$&{\scriptsize{ $\phi^*(z) \simeq \phi^*(z=0)^\mathrm{d}(1+3.17z)$}} \\
&&&&&&\\
\hline
ESS$^\mathrm{e}$    &&&&&&\\
$0.1\la z\la 0.5$  &E+S0+Sa+Sb &436 &$-20.49\,\pm\,0.11$ &$-0.31~\,\pm\,0.14$ &$0.031$& \\
$R_\mathrm{c}\le20.5$ &Sc+Sm/Im            &181 &$-19.84\,\pm\,0.24$ &$-1.63~\,\pm\,0.22$ &$0.006$&{\scriptsize{$\phi^*(z) \simeq \phi^*(z=0.15)^\mathrm{d}[1+3.5(z-0.15)]$}} \\  
&&&&&&\\
\hline
CFRS$^\mathrm{f}$    &&&&&&\\
$0.2\la z\la 1.0$ &Redder than Sbc     &99  &$-20.12\,\pm\,0.25$ &$0.00~~~~~~~~$      &$0.030$& \\
$I\le22.5$    &Bluer than Sbc      &110 &$-20.01\,\pm\,0.25$ &$-1.34~~~~~~~~~~~$  &$0.010$&{\scriptsize{1-mag brightening in $M^*_R$ }}\\
&&&&&& {\scriptsize{modelled here as $M^*_R(z) \simeq$}}\\
&&&&&& {\scriptsize{$M^*_R(z=0.3)^\mathrm{c}-2(1-e^{0.3-z})$}} \\
\hline
LCRS$^\mathrm{g}$   &&&&&&\\
$0\la z\la 0.2$  &clans $1+2+3+4$ &16146 &$-20.42\,\pm\,0.02$ &$-0.13\,\pm\,0.05$  &$0.018$&\\
$r_\mathrm{Gunn}\le17.7$ &clans $5+6$           &2132  &$-20.38\,\pm\,0.08$ &$-1.58\,\pm\,0.07$  &$0.002$ &      \\
&&&&&&\\
\hline
LCRS$^\mathrm{g}$  &&&&&&\\
$0\la z\la 0.2$ &clans $1+2+3$         &12936 &$-20.50\,\pm\,0.03$ &$0.03\,\pm\,0.05$   &$0.012$& \\
$r_\mathrm{Gunn}\le17.7$ &clans $4+5+6$         &5342  &$-20.32\,\pm\,0.06$ &$-1.27\,\pm\,0.05$  &$0.006$ & \\
&&&&&&\\
\hline
1-deg$^2$ mock LZT &&&&&&\\
$0.2\la z\la 1.2$  &early-type  &$\sim$17300 &$-20.42~~~~~~~~~~~$  &$-0.16~~~~~~~~~~~$ &0.03 & \\
$R_\mathrm{c}\le23.0$ &late-type &$\sim$12700 &$-20.18~~~~~~~~~~~$ &$-1.19~~~~~~~~~~~$ &0.01 &{\scriptsize{$M^*_R(z) \simeq M^*_R(z=0.5)^\mathrm{c}$}}\\
&&&&&&{\scriptsize{~~~~~~~~~~~~~~$-2(1-e^{0.5-z})$}} \\
\hline
\end{tabular}
\end{center}
\footnotesize
\underline{Notes:} \\
$^\mathrm{a}$ $\phi^*$ is in units of $h^3$ Mpc$^{-3}$. The quoted uncertainties in $\phi^*$ are 
typically of order of $0.005$. This is a lower limit if one takes into account the fluctuations
caused by the large-scale clustering inside a given survey. For surveys which detect evolution
in $\phi^*$, this Col{.} lists the value $\phi^*(z=0)$ \\
$^\mathrm{b}$ \citet{lin99}; the LFs are measured in the interval $-23\le M_\mathrm{R}\le-17$ \\
$^\mathrm{c}$ value given in Col{.} $M^*_R-5\log h$ \\
$^\mathrm{d}$ value given in Col{.} $\phi^*$  \\
$^\mathrm{e}$ \citet{lapparent02a,lapparent02b}; the LFs are measured in the interval $-23\le M_\mathrm{R}\le-16$\\
$^\mathrm{f}$ \citet{lilly96}; the LFs are measured in the interval $-23.8\le M_B(AB)\le-19$ for red galaxies, and $-23\le M_B(AB)\le-19.5$ for blue galaxies\\
$^\mathrm{g}$ \citet{bromley98}; the LFs are measured in the interval $-23\le M_\mathrm{R}\le-16.5$
\end{table*}

We also show in Table \ref{gallf}, the LFs obtained 
for a different grouping of the LCRS clans: early-type are
galaxies in clans $1+2+3$, and late-type are galaxies in clans $4+5+6$.
The resulting variations in the Schechter luminosity functions
illustrates the sensitivity of the intrinsic LFs to the scheme
used for galaxy classification. The systematically low $\phi^*$ for
both galaxy classes in either groupings ($1+2+3$ and $4+5+6$, $1+2+3+4$ and $5+6$)
as compared to the deeper surveys CNOC2, ESS, and CFRS (see Table \ref{gallf}), 
further suggests that the LCRS suffers from selection effects causing an under-sampling
of the galaxy populations.

The LFs for early and late-type galaxies at $z\ga0.5$ listed in Table \ref{gallf}
show reasonable agreement among the various samples. The dominant
source of variation in the LFs for each galaxy type are caused by the
differences in the definitions of the spectral classes among the
samples. This is reflected in the varying ratio of early to late-type
galaxies among the surveys (see Col{.} $N_\mathrm{gal}$ in Table \ref{gallf}).
\citet{kochanek00} recently showed that the
mixing of the morphological classes which is often present in spectral
classification can cause systematic biases in the parameters of the
LFs. The varying selection effects from survey to survey (such as the 
mentioned bias in the LCRS against low-surface brightness objects)
also contribute to the differences in the LFs.

The deep surveys quoted in Table \ref{gallf} detect evolution of the
LFs with redshift \citep{lilly95,lin99,lapparent02a}.  Understanding the
details of this evolution is still a matter of debate.  Here, we list
some scenarii proposed in the corresponding articles.  In the \citet{lilly95} survey, 
the red galaxies show no or little density
or luminosity evolution in the range $0\la z\la1$, whereas the blue
galaxies show a luminosity evolution of at least 1 magnitude in 
that redshift range. The CNOC2 analysis separates the
luminosity evolution from the density evolution. Early and
intermediate-type galaxies show a small luminosity evolution in the range
$0.1\la z \la 0.7$ ($\Delta M^*\simeq0.5$), whereas late-type galaxies show a
clear density evolution with almost no luminosity evolution in the
same redshift range. Finally, for the ESS, an evolution in $\phi^*$
for the late-type galaxies is detected. These various evolutions are
listed in Table \ref{gallf}, in the Col{.} labeled ``Evolution''.

\begin{figure}
\begin{center}
\epsfysize=6cm
\centerline {\epsfbox[50 50 500 550]{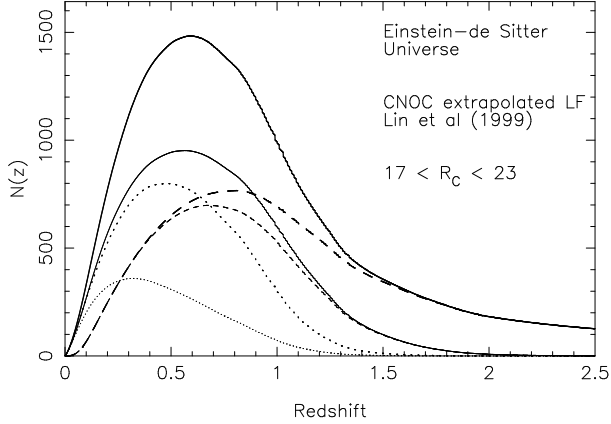}}
\caption{Galaxy redshift distribution in 40 deg$^2$ of the sky to $17\le R_\mathrm{c}\le23$ in an
Einstein-de Sitter Universe, according to the LFs measured from the 
CNOC2 data \protect\citep{lin99} and listed in Table \ref{gallf}. 
The non-evolving and evolving distributions for the early+intermediate galaxies
are shown as a thin dashed line, resp. thick dashed line, and for the 
late-type galaxies, as a thin dotted line, resp. thick dotted line. 
The non-evolving and evolving total distributions are shown as thin solid line, 
resp. thick solid line.}
\label{cnocnz}
\end{center}
\end{figure}

\begin{figure}
\begin{center}
\epsfysize=6cm
\centerline {\epsfbox[50 50 500 550]{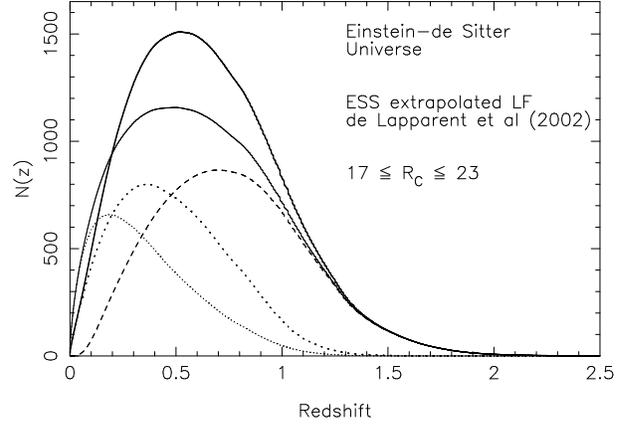}}
\caption{Galaxy redshift distribution in 40 deg$^2$ of the sky to $17\le R_\mathrm{c}\le23$ in an
Einstein-de Sitter Universe, according to the LFs measured from the 
ESS data \protect\citep{lapparent02a,lapparent02b} and listed in Table \ref{gallf}.
The non-evolving distribution for the early-type galaxies is shown as a thin 
dashed line, and for the late-type galaxies, as a thin dotted line. The evolving 
distribution for the late-type galaxies is shown as a thick dotted line, and is 
modeled as $\phi^*(z) \simeq \phi^*(z=0.15)[1+3.5(z-0.15)]$. The non-evolving and evolving 
total distributions are shown as thin solid line, resp. thick solid line.}
\label{essnz}
\end{center}
\end{figure}

\begin{figure}
\begin{center}
\epsfysize=6cm
\centerline {\epsfbox[50 50 500 550]{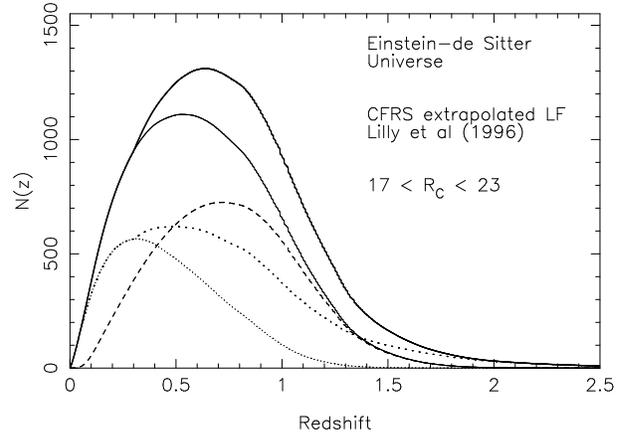}}
\caption{Galaxy redshift distribution in 40 deg$^2$ of the sky to $17\le R_\mathrm{c}\le23$ in an
Einstein-de Sitter Universe, according to the LFs measured from the 
CFRS data \protect\citep{lilly95} and listed in Table \ref{gallf}. 
The non-evolving distribution for the red galaxies is shown as a thin dashed line,
and for the blue galaxies, as a thin dotted line. The evolving distribution for 
the blue galaxies is shown as a thick dotted line, and is modeled as
$M^*_R(z) \simeq M^*_R(z=0.3)-2(1-e^{0.3-z})$ (see Table \ref{gallf}).
The non-evolving and evolving total distributions are shown as thin solid line, 
resp. thick solid line.}
\label{cfrsnz}
\end{center}
\end{figure}

\begin{figure}
\begin{center}
\epsfysize=6cm
\centerline {\epsfbox[50 50 500 550]{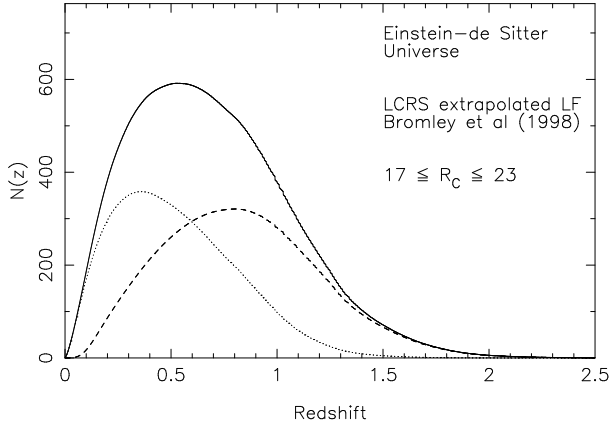}}
\caption{Galaxy redshift distribution in 40 deg$^2$ of the sky to $17\le R_\mathrm{c}\le23$ in an
Einstein-de Sitter Universe, according to the LFs measured from the 
LCRS data \protect\citep{bromley98} and listed in Table \ref{gallf}.
The distribution for the galaxies in clans $1+2+3$ is shown as a 
thin dashed line, and as a thin dotted line for galaxies in clans $4+5+6$.
The thin solid line shows the total distribution.}
\label{lcrsnz}
\end{center}
\end{figure}

Figures \ref{cnocnz}, \ref{essnz}, and \ref{cfrsnz} show for
each survey displayed in Table \ref{gallf} the redshift distributions
calculated for the listed Schechter parameters: without evolution
(thin dashed line for early-type galaxies, and thin dotted 
line for late-type galaxies), and with evolution when it applies
(thick dashed line for early-type galaxies, thick dotted line for late-type
galaxies). These predicted redshift distributions are calculated in the
case of a flat Universe with $\Lambda=0$ and $H_0 = 100 h$ km s$^{-1}$
Mpc$^{-1}$, over the $40$ deg$^2$ planned area for the LZT survey, and
are extrapolated to the limiting magnitude of the LZT
survey, namely $R_\mathrm{c}\le23.0$.  In Fig{.} \ref{cfrsnz}, we model the
observed evolution of the CFRS blue LF with a brightening in $M^*_R$, defined by
the additive term $m(z)=-2[1-\exp -(z-0.3)]$; this yields $m=0.0$
for $z=0.3$, $m=-1.0$ for $z=0.6$, $m=-1.9$ for $z=1.0$, and $m=-2.7$
for $z=1.5$. For comparison, the redshift distributions for the
early and late-type LFs for the LCRS divided in clans $1+2+3$ and $4+5+6$ 
are shown in Fig{.} \ref{lcrsnz} (no evolution is considered). 

In Figs{.} \ref{cnocnz}, \ref{essnz}, \ref{cfrsnz}, 
the systematic differences between the evolving early-type/red and late-type/blue galaxy redshift
distributions show common properties. The peak for the late-type distribution is at smaller
redshift than that for the early-type distribution because of the combination
of fainter $M^*$ and steeper slope $\alpha$ for the first class of objects. 
This effect is preserved when evolution of the early-type and/or late-type 
population are introduced, despite the fact that all evolution scenarii quoted 
in Table \ref{gallf} tend to shift the peaks of the redshift distributions
to higher redshift.
The interesting result derived from the comparison of Figs{.}
\ref{cnocnz}, \ref{essnz}, \ref{cfrsnz}, is that the 3
parameterizations of the LF evolution lead to resembling galaxy redshift
distributions in the range $0\la z\la 1$ at the depth of the LZT
survey ($R_\mathrm{c}\le23.0$): the redshift and amplitude at the
peak, as well as the high redshift fall-off of the redshift distributions
are similar. In contrast, the LCRS produces a systematically low redshift distribution
at this depth, with a peak at $\sim600$ galaxies, whereas the
no-evolution curves for the deeper surveys shown in Figs{.} \ref{cnocnz}, \ref{essnz},
\ref{cfrsnz}, all peak in the range $\sim900-1200$ galaxies; as already mentioned,
this may arise from selection effects in the LCRS.

We therefore choose to adopt for all mock catalogues considered here
the following LF Schechter parameters: $M^*_R=-20.42$, $\alpha=-0.16$,
$\phi^*=0.03$ for early-type galaxies; $M^*_R=-20.18$,
$\alpha=-1.19$, $\phi^*=0.01$ for late-type galaxies (also listed in Table \ref{gallf}). 
These chosen $\alpha$ for the early-type galaxies, and the $M^*_R$ and $\phi^*$ for
both galaxy types are within the range of values measured for the 3 $R$-selected 
surveys in Table \ref{gallf}: the CNOC2, the ESS, and the LCRS. 
The slope $\alpha=-1.19$ for the late-type galaxies is flatter than
the flattest measured slope (from the LCRS, clans $4+5+6$),
in order to limit the number of galaxies to be included in the mock LZT catalogues, 
and thus the computing time for the PCA;
this value is however only $1.25\sigma$ flatter 
than the slope of the LF for the CNOC2 late-type galaxies.
Having a steeper slope for the late-type galaxies would have a weak impact on the
results presented here.
Note that the chosen values of $\phi^*$ for the early and late-type LFs
are obtained by using an early-type to late-type $\phi^*$ ratio of 
3 (as in the CFRS) and by normalizing the total number 
of galaxies per deg$^2$ with $R_\mathrm{c}\le23$ to the observations: numerous
studies give galaxy counts in the $R$ band, and we use a typical value
of $\sim30,000$ galaxies/deg$^2$ \citep{roche96,metcalfe01}.

\begin{figure}
\begin{center}
\epsfysize=6cm
\centerline {\epsfbox[50 50 500 550]{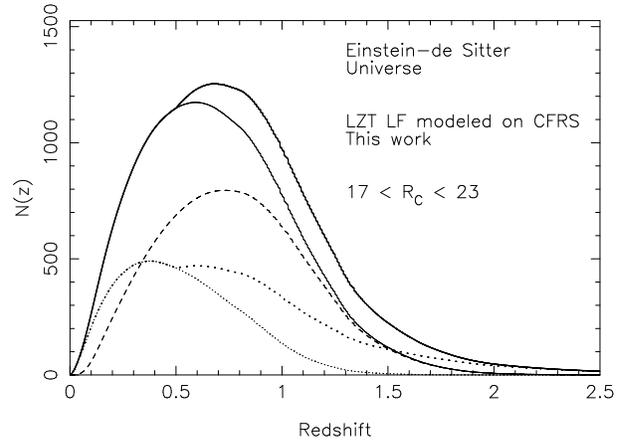}}
\caption{Predicted redshift distribution for the future LZT survey in 40 deg$^2$ 
of the sky to $17\le R_\mathrm{c}\le23$ in an Einstein-de Sitter Universe. The redshift 
distributions for the non-evolving early-type, late-type galaxies are shown as a 
thin dashed line, resp. thin dotted line.
The corresponding Schechter parameters are $M^*_R=-20.42$, $\alpha=-0.16$, 
$\phi^*=0.03$ for the early-type galaxies, and $M^*_R(z\le0.5)=-20.18$, 
$\alpha=-1.19$, $\phi^*=0.01$, for the late-type galaxies. The expected 
distribution for the evolving late-type galaxies, modeled by an $M^*_R$ 
brightening define by $M^*_R(z\ge0.5) \simeq M^*_R(z=0.5)-2(1-e^{0.5-z})$ is 
shown as a thick dotted line. The thin, resp. thick continuous lines show the 
sums of the curves for the non-evolving early-type galaxies and the non-evolving, 
resp. evolving late-type galaxies. The evolving sum is subsequently used 
for generating the mock LZT catalogues.}
\label{totnz}
\end{center}
\end{figure}

Because the LZT survey will have
a depth comparable with that of the CFRS, we adopt an evolution of the LF
resembling that for the CFRS, but which better matches the
evolution at $z>1$ (as shown in Fig{.} \ref{totnz}, a non negligeable
fraction of galaxies, $27$\%, will have $z\ge1$ in the LZT
survey). Although the evolution of the galaxy LFs is poorly known beyond
$z=1$, photometric redshifts applied to the Hubble Deep Field do
provide a general trend for the evolution of the ``total'' LF: a
steepening of the slope $\alpha$ from $-1.3$ ($z\sim 0.5$) to $-2$
($z\sim 3$), and a one magnitude brightening of $M^*_B$ between $z\sim
1$ and $z\sim 3$ \citep{sawicki97,takeuchi00}.  We therefore conservatively assume
no evolution of both the early and late-type galaxy LFs in the range
$0\le z\le0.5$, and add a brightening term $m(z)=-2[1-\exp -(z-0.5)]$ to $M^*_R$ 
for the LZT blue galaxies with $z\ge0.5$.  The brightening term $m(z)$
gradually changes $M^*_R$ by $m=-0.83$ at $z=1$, $m=-1.0$ at $z=1.19$,
$m=-1.26$ at $z=-1.5$, $m=-1.55$ at $z=2$, and $m=-1.8$ at $z=3$; it
asymptotically reaches its ceiling of $m=-2$ at $z\ge3$. 
The corresponding redshift distribution for the evolving late-type galaxies 
in the modeled LZT survey is given in Fig{.} \ref{totnz} (thick dotted line);
it is combined with the non-evolving early-type redshift distribution (thin dashed line) 
to derive the sum of the 2 populations in the evolving model (thick continuous 
line). This summed redshift distribution is subsequently adopted for the LZT 
mock surveys analyzed here. For comparison, the corresponding non-evolving 
late-type galaxies distribution (thin dotted line) and non-evolving total 
distribution (thick continuous line) are also shown in Fig{.} \ref{totnz}.

\begin{figure*}
\begin{center}
\epsfysize=19cm
\centerline {\epsfbox[35 23 545 733]{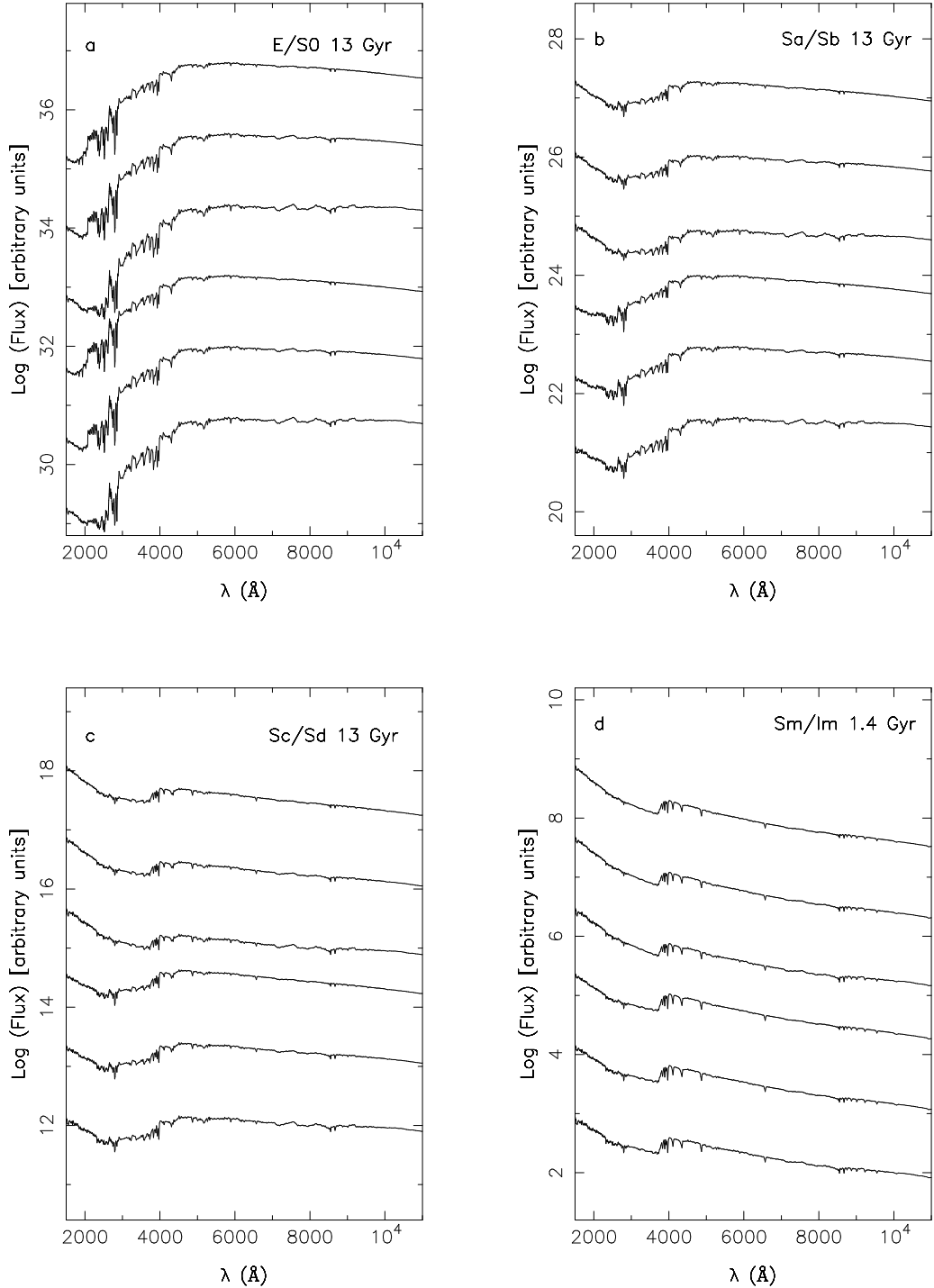}}
\caption{Galaxy template spectra from GISSEL \protect\citep{bruzual93}.
In each frame, the Initial Mass Function (IMF) and metallicity of the
spectra vary from top to bottom as follows: Salpeter IMF, with 20\% solar, 40\% solar, and solar metallicity; 
Scalo IMF, with 20\% solar, 40\% solar, and solar metallicity.}
\label{galgissel}
\end{center}
\end{figure*}

\begin{figure*}
\begin{center}
\epsfysize=13cm
\centerline {\epsfbox[50 0 500 550]{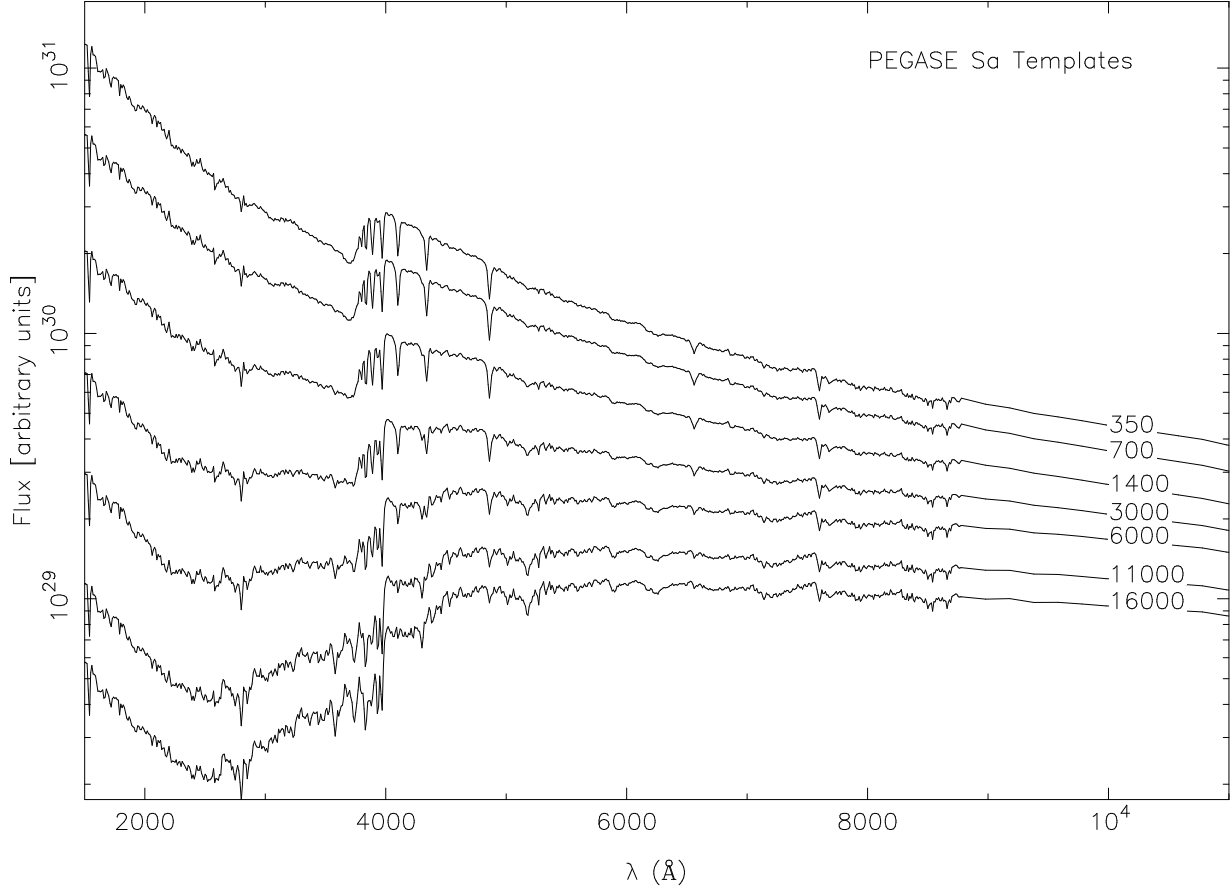}}
\caption{Sa13 galaxy template spectra from PEGASE \protect\citep{fioc97}.
The age of the spectra are given in million years. The metallicity is adjusted 
to have a solar value for the 13-Gyr-old spectrum.}
\label{galpegase}
\end{center}
\end{figure*}

Over the planned area of $40$ deg$^2$ for the LZT survey, over 
$\sim1,000,000$ galaxies are expected. Because the PCA is computer-time 
consuming, we limit the size of the LZT mock catalogues used in the 
following analyses to 1 deg$^2$. These calatogue therefore include $\sim30,000$ 
galaxies, and the number of galaxies in each class are those listed in 
Table \ref{gallf}.

\subsubsection{Galaxy spectral energy distributions}

The galaxy templates are extracted from stellar synthesis libraries. Figure
\ref{galgissel} shows 24 templates of E/S0, Sa/Sb, Sc/Sd, and Sm/Im galaxies
for 3 metallicities and 2 Initial Mass Functions (IMF) kindly
provided by S. Charlot, and calculated from the version GISSEL95
of the GISSEL evolutionary code \citep{bruzual93}.  For early-type
galaxies, higher metallicities flatten the slope of the
continua. Changing the IMFs induces no effect on the templates (Fig{.}
\ref{galgissel}, frame a). For late-type galaxies, the Salpeter IMF \citep{salpeter55} 
tends to produce bluer objects than the Scalo IMF \citep{miller79}, and the 
metallicity effect is small
(Fig{.} \ref{galgissel}, frames b to d). The Sa/Sb templates using a Salpeter IMF 
are very similar to the Sc/Sd templates using a Scalo IMF, and at the resolution of 
the LZT survey, they are nearly identical.

Alternatively, the PEGASE package \citep{fioc97} allows us to generate a set of solar 
metallicity spectra with different ages, different stellar formation rates (SFR) and
different IMF taken from \citet{rana92}.
Details can be found in Fioc \& Rocca-Volmerange (1997).
Included in the PEGASE package is an atlas of templates of eight galaxy types.
For the various spectral types (E, S0, Sa, etc...), the atlas provides 65 templates 
from {\emph{ages}} of 1 Myr to 16 Gyr;
each age sequence is normalized so that the 13-Gyr templates fit 
present day SEDs of observed galaxies with solar metallicity.

Even if true spectra cover a large range of metallicities and ages, the resolution
$R\simeq40$ of the LZT survey does not allow us to distinguish 
between a PEGASE {\emph{young}} Sa galaxy template and PEGASE {\emph{old}} Sb, Sc, Sd or 
Im galaxy template. Therefore, we include in the mock LZT catalog only templates 
showing {\emph{significant}} differences and instead of using the {\emph{ages}} 
given by PEGASE as true indicators of the evolution of galaxies with redshift,
we use them as indicators of the morphological sequence.
Hence, for elliptical galaxy SEDs, we use the 23 templates of the PEGASE atlas 
called E13 and older than 1 Gyr (the templates have a short SFR, followed by passive evolution).
For spiral galaxy SEDs, we use the 31 templates of the PEGASE atlas 
called Sa13, older than 350 Myr (we use a constant SFR for spiral galaxies).
Figure \ref{galpegase} shows several Sa13 templates from the PEGASE library, shown at various 
{\emph{ages}}. As already mentioned, the 13-Gyr-old template is fitted to a local
Sa galaxy spectra with solar metallicity. We interpret the {\emph{ages}
sequence given by PEGASE Sa13 templates as follows:
late-type galaxies (Sc, Sd, Im) are modeled by {\emph{young}} Sa13 galaxy 
templates (with ages $<3$ Gyr, 15 templates), and early-type galaxies (Sa, Sb) by 
{\emph{old}} Sa13 galaxy templates (with ages $\ge3$ Gyr, 16 templates).

There are notable differences between PEGASE and GISSEL elliptical galaxies
in the UV, whereas spiral templates look alike in both models.
We also emphasize that both GISSEL and PEGASE packages rely on the stellar
library of Pickles which is noticeably deficient in AGB stars SEDs. AGB
stars are still poorly known but represent a non-negligeable part of the
total integrated flux of galaxies.

The mock LZT galaxy catalogues are built using the (evolving) redshift distributions 
in Fig{.} \ref{totnz}. 
Early-type galaxies are randomly extracted from the 12 GISSEL E/S0/Sa/Sb templates,
the 23 PEGASE E13 templates, and the 16 Sa13 templates older than 3 Gyr.
Late-type galaxies are randomly extracted from the 12 GISSEL Sc/Sd/Sm/Im
templates, and the 15 PEGASE Sa13 templates younger than 3 Gyr. 
The number of early and late-type galaxies are those 
plotted in Fig{.} \ref{totnz} as non-evolving early-type
and evolving late-type redshift distributions for the LZT survey (see also
Table \ref{gallf}).

\subsubsection{Emission-line galaxies}
We do {\em not} model the emission lines in the galaxy SEDs,
because in the medium-band filters used by the LZT, only the brightest
emission lines will be detected. At first order, an emission line
will be detected only if 
\begin{equation} \frac{W_\mathrm{line}}{W_\mathrm{filter}} \ge \frac{Threshold}{S/N},
\label{eqline}\end{equation}
where $W_\mathrm{line}$ is the line equivalent width, $W_\mathrm{filter}$ is the filter
bandwidth, $Threshold$ is the detection threshold in units of standard
deviation, and $S/N$ is the signal-to-noise ratio. For $Threshold =1$, 
$S/N=3$, and a typical $W_\mathrm{filter}=150$\AA, the emission line must
have $W_\mathrm{line}\ge$ 50\AA. Only QSOs and Seyfert galaxies reach this level of 
emission. 

We emphasize that the analysis of the QSO sub-sample included 
in the LZT mock catalog (see next Sect{.}) demonstrates that strong
emission lines contribute to improving the object classification and
the redshift determination which are obtained by the PCA. As seen in
Fig{.} \ref{plotpcamethod} and described in Sect{.} 3.1 below, the
PCA sequence of QSOs is clearly separated from the blue part of the
stellar sequence, and this is caused by the strong emission lines
present in the QSO SEDs. 

Although Seyfert galaxies only represent a
small fraction ($5\%$) of the galaxy populations at low redshift
\citep{reichert92}, the fraction of galaxies with strong emission lines
may be larger at $z\ga0.5$. However, we consider that it is not
necessary to include them in the mock LZT catalogues, as they have
similar SEDs to QSOs, and would therefore make a negligeable change to
the PCA eigenbasis. Seyfert galaxies and other galaxies with strong
emission lines would deviate from the locus of normal galaxies in the
PCA, and would thus be easily identified; their redshift would be
measured with similar accuracy as for the QSOs (see Sect{.}
\ref{redshift}). They could also be directly identified from their SEDs
using a break-finding algorithm \citep{cabanac98}, or a
cross-correlation analysis (as shown by preliminary tests performed by
R. Cabanac).

\subsection{Quasi-stellar objects}\label{QSOs}

The third kind of objects included in the simulations are the quasi-stellar
objects (QSO). We use the preliminary results of the on-going 2dF QSO
Survey \citep{boyle00}. The 2dF QSO survey has been optically selected in the $U$, $B_J$ and $R$ 
bands from UKST photographic plates. The $B_J$ QSO LF is found to follow a pure luminosity
evolution \citep{boyle00} which can be modeled by
\begin{eqnarray}
\phi(L_B,z)dL&=& \frac{\phi_B^*dL}{\displaystyle\left[\frac{L_B}{L_B^*(z)}\right]^{3.6}+\left[\frac{L_B}{L_B^*(z)}\right]^{1.8}}\\
\nonumber{\rm with~~~~~~~~~~~~~~}\\
\nonumber\\
\log\left[\frac{L_B^*(z)}{L_B^*(0)}\right]&=&1.4z-0.27z^2,\\
\nonumber\\
{\rm or\,\, equivalently}\nonumber\\
M_B^*(z) &=& M_B^*(0) -2.5\,(1.4z-0.27z^2),
\end{eqnarray}
We assume the same analytical description for the $R$ QSO LF, and
choose to adopt for the simulations $M_R^*\simeq-21$ and $\phi_R^*=10^{-6}$ Mpc$^{-3}$ mag$^{-1}$,
a $\sim 35$\% larger value than calculated from the $B_J$ 2dF survey \citep{boyle00}.
Equation 3 can be rewritten in terms of absolute magnitude $M_R$:
\begin{eqnarray}
\phi(M_R,z)dM_R&=&\frac{0.4ln(10)\Phi_R^*dM_R}{X^{3.6-1}+X^{1.8-1}},\nonumber\\
X &=& 10^{0.4[M^*_R(0)-M_R]-1.4z+0.27z^2}. \label{qsoeqlf}\end{eqnarray}
We assume that Eq{.} \ref{qsoeqlf} is a reasonable prediction of the LZT QSO
sample and we extrapolate the QSO LF to the LZT apparent magnitude limit of $R_\mathrm{c}\le23$.
We also set a bright cut off at $R_\mathrm{c}=15$. The resulting redshift distribution
in the interval $0\le z\le4$ is shown in Fig{.} \ref{qsonz}.

\begin{figure}
\begin{center}
\epsfysize=6cm
\centerline {\epsfbox[50 50 500 550]{MS1340f10.eps}}
\caption{Redshift distribution for QSOs with $15\le R_\mathrm{c}\le23$
in an Einstein-de Sitter Universe, using the 2dF QSO survey LF \protect\citep{boyle00}.}
\label{qsonz}
\label{qso}
\end{center}
\end{figure}

\begin{figure}
\begin{center}
\epsfysize=6cm
\centerline {\epsfbox[50 50 500 550]{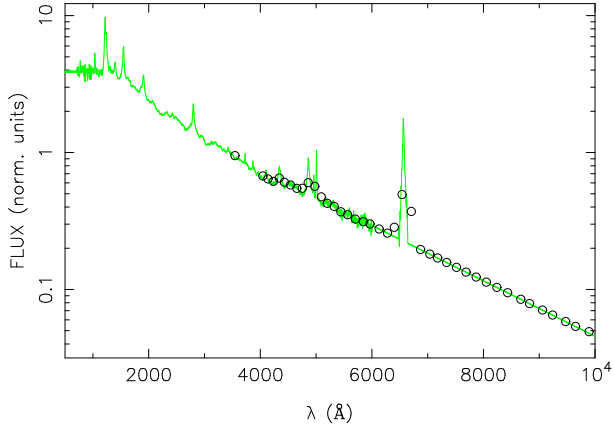}}
\caption{Composite QSO spectrum at $z=0$ \protect\citep{francis91}. 
The wavelengths of the centers of the LZT filters are overlaid ($\circ$). At red wavelengths,
the spectrum is extrapolated by a power-law from 6000 to 10$^4$ \AA, to which is added
a synthetic H$_\alpha$ line. In the UV, the spectrum is completed 
with observations from HST \protect\citep{zheng97}.}
\end{center}
\end{figure}

Figure \ref{qso} shows a composite QSO template derived from 
the LBQS sample \citep{francis91}, and a sample of 101 QSOs observed in the UV
using HST \citep{zheng97}. Because the QSO template provided by Francis et
al. is bounded at 6000 \AA, and the LZT filters extend out to 10$^4$ \AA, we approximate the
missing continuum by a power-law in the range 6000-10 000 \AA. In the blue, the 
composite spectrum is bounded at 800 \AA. In order to define the U and blue LZT magnitudes 
for all QSOs with $z\le4$, we extrapolate the spectrum to 300 \AA\ at
a constant flux. Note that because of Lyman $\alpha$ absorption, the continuum of the
spectrum is poorly defined at wavelengths bluer than $\sim 1200$ \AA\ \citep[see][]{vandenberk01}; 
this choice only affects QSOs with $z\ge 2.75$, among which,
only those with weak emission lines would be ``lost'' in the star sample. We also add a
synthetic $H_\alpha$ emission line whose intensity is related to the 
intensity of the $H_\beta$ emission line already present in the composite spectrum,
according to the typical $H_\alpha/H_\beta$ line ratio of $\sim4$,
obtained in models of broad line regions \citep{osterbrock89}.
A new composite spectra has been made available by the SDSS consortium
\citep{vandenberk01} while the present article was in the refereeing process.
Because both the emission lines and the continua of the 2 composite
spectra are remarkably similar, we did not update our simulations to
include the SDSS composite spectrum: the wide spectral range
of the SDSS spectrum would make no improvement to our analysis, as the
range covered by the 2dF composite spectrum used here is sufficient to
describe the full wavelength interval of the LZT filters for QSOs with
$0\la z\la 4$.

The integrated QSO number count per deg$^2$ is extrapolated from the
differential number counts in the 2dF QSO survey \citep{boyle00}. To a limiting
apparent magnitude of $R\le23$, the integrated count of QSOs per square degree
is $\sim130$. The mock LZT catalogues are generated for 1 deg$^2$, hence
each simulation is obtained by randomly drawing $130$ QSO templates with 
the same redshift distribution as in Fig{.} \ref{qsonz}, and subsequently adding 
noise according to the LZT efficiency (see the next Sect{.}). 
The range of signal-to-noise ratio 
in the continuum of the synthetic QSO spectra is $3-100$.

Because the number of QSOs is relatively small in our mock catalogues
we decided to use only the composite SED of Francis without slope variations.
Nevertheless, it is likely that the slopes of the continuum of real QSO SEDs vary
around the mean slope of the composite SED of Francis. The effect of such a
variation on the PCA would be to spread the locus of QSOs from a line to a 
surface; the area of this surface would be related to the standard deviation in the variation 
of the slope. This variation could degrade the identification of QSOs, 
because the flatter the slopes, the higher the similarities between 
QSOs and emission-line galaxies. On the other hand, the measured redshift 
accuracies (Sect{.} 5) should not be affected by this simplification.

\subsection{Noise}\label{noise}

Photon noise and detector read-out noise are added to the SED of each type 
of object according to the response curve $T(\lambda)$ of the Large 
Zenith Telescope (LZT), defined as the product of the detector sensitivity curve
$CCD(\lambda)$ by the transmission curve of the telescope/instrument optics $O(\lambda)$,
and the sky transmission $I(\lambda)$:
\begin{equation}
T(\lambda) = I(\lambda)\; O(\lambda)\; CCD(\lambda);
\end{equation}
For an object with intrinsic flux $F_0(\lambda)$, the final flux $F(\lambda)$
obtained after ``observing'' the object with the LZT telescope+instrument+detector
and correcting it using an absolute flux calibration is 
\begin{equation}
F(\lambda)= F_0(\lambda) \left[1+\frac{gauss}{S/N(\lambda)}\right]
\end{equation}
with
\begin{eqnarray}
S/N(\lambda) &=& \frac{\sqrt{\frac{T(\lambda)\;N_\mathrm{night}}{ g(\lambda)}}\; F_0(\lambda) }
	              {\sqrt{F_0(\lambda)+A_\mathrm{src}\; sky(\lambda)+
                             \frac{A_\mathrm{src}\; RON^2\; g(\lambda)}{A_\mathrm{pix}\; T(\lambda)\;N_\mathrm{night}}}},\label{sn}\\
g(\lambda) &=& \frac{10^{20}\; hc}{\lambda_c\;\Delta\lambda\;\Delta t\; A_\mathrm{mirror}}.
\label{g_lambda}
\end{eqnarray}
$gauss$ is a gaussian random generator, with a null mean and a
root-mean-square dispersion of 1, $S/N(\lambda)$ is the ``observed''
signal-to-noise ratio in the spectrum $F(\lambda)$; $sky(\lambda)$
is a composite sky spectrum, using Kitt Peak night sky spectrum
\citep{massey00}, Mauna Kea KECK LRIS OH emission lines atlas
(http://www2.keck.hawaii.edu), and GEMINI near IR modeled continuum
(http://www.us-gemini.noao.edu). Figure \ref{plotsky} shows the
composite high-resolution sky spectrum and its medium-band counterpart
using the LZT filters.  In Eq{.} \ref{sn}, $A_\mathrm{src}$ is the area of the observed object
in arcsec$^2$ on the detector, $A_\mathrm{pix}=0.245$ arcsec$^2$ is the area
of one CCD pixel, $RON=11$ e$^-$ is the CCD readout noise.  The
function $g(\lambda)$ in Eq{.} \ref{g_lambda} is the flux expressed in Joule/second/meter$^3$
corresponding to one photon having the central wavelength $\lambda_c$
of the LZT filter considered (see Fig{.} \ref{transcurv}), 
arriving on the detector per exposure time $\Delta t$ and per wavelength 
interval $\Delta\lambda$ (of the considered filter),
given the area of the LZT mirror $A_\mathrm{mirror}=28.3$ m$^2$ (wavelengths are
expressed in meters, time in seconds); $hc=1.992\time10^{-25}$ J$\cdot$m is the
product of the Planck constant by the speed of light.  Because the LZT 
is operated in drift-scan
mode, the exposure time is fixed to $\Delta t=65$ sec. The exposure
time is effectively increased by observing a given region of the
survey in a given filter during several nights $N_\mathrm{night}$ (see Eq{.} 9).

\begin{figure}
\begin{center}
\epsfysize=6cm
\centerline {\epsfbox[50 50 500 550]{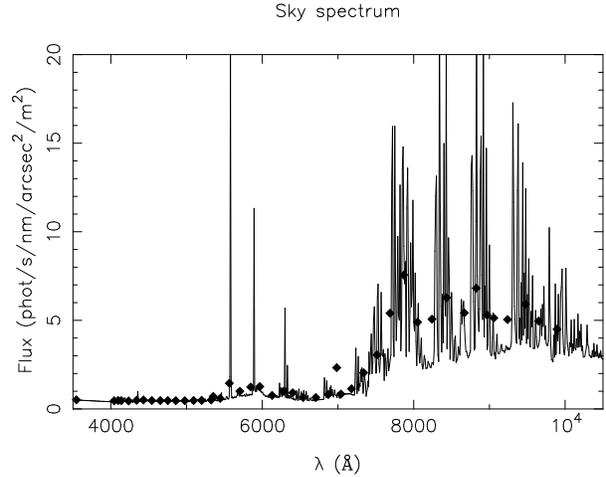}}
\caption{Composite sky spectrum using Kitt Peak night sky spectrum
\protect\citep{massey00}, Mauna Kea KECK LRIS OH emission lines atlas
(http://www2.keck.hawaii.edu), and GEMINI near IR modeled continuum
(http://www.us-gemini.noao.edu). The filled diamonds are the LZT
U+medium-band filter fluxes used for noise computations.  }
\label{plotsky}
\end{center}
\end{figure}

\begin{figure}
\begin{center}
\epsfysize=12cm
\centerline {\epsfbox[100 0 500 750]{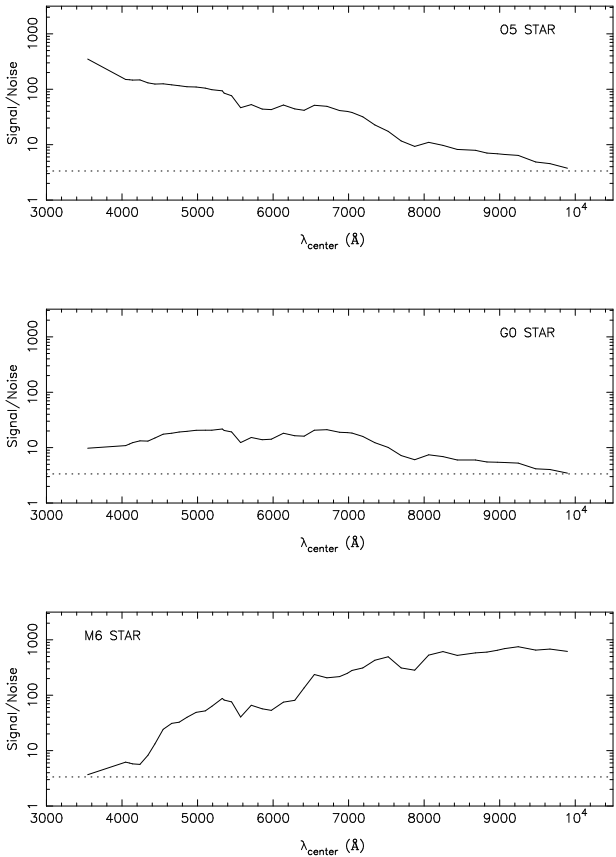}}
\caption{Signal-to-noise distributions ($S/N[\lambda]$) of 3
stellar spectra with types O5V, G0V, and M6V, with median $S/N$ of 44,
14, and 122 resp{.}. The lowest $S/N$ is set to $3.35$ (dotted line):
its corresponds to the signal-to-noise ratio of a punctual source extended
over 5 contiguous pixels, with a detection threshold of 1.5 per pixel.}
\label{spectnoise}
\end{center}
\end{figure}

The number of optical surfaces in the telescope/instrument combination
is 11: the mirror (80\% reflectivity), and 4 lenses plus the CCD window
(98\% transmission is assumed for each of the 10 surfaces). For simplicity the 
transmission is supposed to be achromatic.
The LZT total efficiency curve $T(\lambda)$ has a bell-shape between
3000\AA\, and 10000\AA, with a flat maximum transmission of 0.65
running from 5000\AA\, to 8000\AA. Figure \ref{spectnoise} shows the
$S/N(\lambda)$ calculated using Eq{.} \ref{sn} for an O5V star, a G0V
star and an M6V star, with their lowest signal-to-noise ratio set to
3.35: this corresponds to the overall signal-to-noise ratio obtained
from 5 contiguous pixels having each a signal-to-noise ratio of 1.5.
Note that we choose such a low detection level per pixel 
in order to test the PCA.  Given the LZT median seeing of 0.9 arcsec
FWHM (Full-Width-Half-Maximum), we do not expect stars to be less
extended than 5 pixels, as it corresponds to the area of a disk of
1.2-arcsec diameter projected onto CCD pixels with an area
$A_\mathrm{pix}=0.245$ arcsec$^2$. We thus adopt the limiting signal-to-noise
ratio of $3.35$ as our detection threshold for the LZT mock
catalog. For galaxies at $z\simeq0.5$, a typical FWHM of 3 arcsec
(see Sect{.} \ref{extended}) yields an area of 28.8 pixels, and a
detection threshold of $S/N=1.5$ per pixel yields an overall $S/N\sim
8$. 

We also define the overall signal-to-noise of each spectrum as the
\emph{median} value of $S/N(\lambda)$ over the 41 LZT filters. For the
spectra in fig{.}  \ref{spectnoise}, these median $S/N$ are 44 (O5V),
14 (G0V), and 122 (M6V). Figure \ref{spectnoise} therefore illustrates that
very blue or very red spectra must have a high median $S/N$ 
to be detected over the entire range of the LZT filters.
To simplify the analysis and interpretation, the SEDs of all
classes/types of objects in a given mock catalog are set to the \emph{same}
median signal-to-noise ratio. Mock catalogues are generated such
that only objects with a lower filter above the threshold limit of
$S/N(\lambda)=3.35$ and median signal-to-noise ratio equal to a given
value are included: any object having at least one filter with a
$S/N(\lambda)<3.35$ is therefore not included in the catalog. 
Catalogues with median signal-to-noise ratios of $100,\,20,\,10$, 
and $6$ are used for the analysis reported here.  In the
following Sects{.}, the labels $S/N$ always refer to the \emph{median} 
signal-to-noise ratio of the spectra in the mock catalogue
considered. As shown in Fig{.} \ref{spectnoise}, we emphasize that 
at a median $S/N$ ratio of 100, the bluest objects will have
a signal-to-noise ratio of $\sim10$ in their reddest filters,
whereas the reddest objects will have a signal-to-noise ratio of 
$\sim3$ in their bluest filters; for flat-spectrum objects,
the range of signal-to-noise ratio described by the spectra
will be narrower.

\section{Methodology}\label{method}

\subsection{Principal Component Analysis}

For a detailed description of the PCA, the reader is referred to the books
of \citet{murtagh87} or \citet{kendall72},
and to the seminal paper of \citet{connolly95b}. In this section we
outline the results of PCA and develop the classification method.\\
Following \citet{connolly95b}, consider a set $f$ of $N$ vectors 
$f_{i}$ of $M$ elements $f_{\lambda i}, ~i\in\{1\ldots N\},~\lambda\in\{1\ldots M\}$, 
normalized to have unit scalar products $[f_{i}\cdot f_{i}]^{1/2} = 1$.
In our case, $M$ is the number of filters: $M=41$. The PCA derives a set 
of $min(N,M)$ orthogonal eigenvectors $e_j$ (that is $M=41$ eigenvectors in the
present case, as $N\gg M$), using criteria of decreasing 
maximum variance of the spectra when projected onto the eigenvectors. 
Each vector $f_{i}$ can be written as a linear combination of $e_j$:\\
\begin{equation}
  f_{\lambda i} = \sum_{j=1}^M y_{ij}e_{\lambda j},\label{eqM}
\end{equation}
where $y_{ij}$, denoted eigencomponent, is the weight of the $j$th eigenvector 
in the $i$th vector.
The first eigenvector $e_1$ is the mean vector over the $f_{\lambda i}$. Each weight $y_{i1}$ measures
how much $f_{\lambda i}$ is similar to the mean vector, i.e. gives
its projection onto the mean vector; the second vector $e_2$ lies in the
direction of highest variance orthogonal to $e_1$ etc. 

The main advantage of the PCA is that when the vectors $f_{i}$ are
correlated (as it is the case for astronomical SEDs), most of the discriminatory 
power of the linear combination (Eq{.} \ref{eqM}) is carried by the first few eigenvectors,
and the high-order eigenvectors carry mostly the noise of the spectra.
The PCA therefore provides a powerful filter for the set $f$ (see Galaz
and de Lapparent, 1998).
For illustration, Table \ref{tablemethod} shows the eigenvalues
$\gamma_j$ of a PCA performed on one mock LZT catalog described in Sect{.} 2.
As shown by \citet{connolly95b}, each eigenvalue $\gamma_j$ is the contribution of the corresponding 
eigenvector $e_j$ to the variance of the set $f$, and therefore describes the relative
power of each eigenvector in the dataset (one can intuitively realize that if all vectors 
have similar directions, they can be described by a small number of components).
The power $P_j$ carried by each eigenvector $e_j$ can be measured as
\begin{equation}
P_j = \frac{\gamma_j}{\sum_{j=1}^M \gamma_j}.\label{eqgamma}
\end{equation}
Table \ref{tablemethod} therefore shows
that the first 3 eigenvectors $e_1$, $e_2$, $e_3$ 
carry 87.6\%, 9\%, 2.2\% resp{.} of the descriptive power; the sum of these 3 contributions
yields 98.8\%, and adding $e_4$ increases the summed contributions to 99.3\%.
This means that the first 3 vectors actually
carry most of the information and, to a good approximation, the weights of the
other eigenvectors may be neglected. This is the reason why only the first
3 or 4 eigencomponents are usually kept to describe a set of galaxies at
$z=0$ \citep{ronen99}. 

\begin{table}
\begin{center}
\caption{Eigenvalues $\gamma_j$ of a typical PCA derived eigenbasis $e$
from a LZT mock catalog 
which contains a mix of 3368 stars, 11600 galaxies and 170 QSO SEDs with $S/N=100$, 
as described in Sect{.} 2. The first 3 eigenvalues
contain 98.8\% of the descriptive power.}
\label{tablemethod}
\begin{tabular}{cr|cr|cr}
\hline
\hline
$e_j$&$\gamma_j$&$e_j$&$\gamma_j$&$e_j$&$\gamma_j$\\
\hline
$e_1$&13230.44&$e_{15}$&1.198&$e_{29}$&0.076\\
$e_2$&1355.23&$e_{16}$&1.004&$e_{30}$&0.070\\
$e_3$&340.38&$e_{17}$&0.972&$e_{31}$&0.057\\
$e_4$&71.54&$e_{18}$&0.775&$e_{32}$&0.055\\
$e_5$&35.57&$e_{19}$&0.703&$e_{33}$&0.051\\
$e_6$&25.90&$e_{20}$&0.563&$e_{34}$&0.007\\
$e_7$&16.70&$e_{21}$&0.465&$e_{35}$&0.010\\
$e_8$&6.434&$e_{22}$&0.426&$e_{36}$&0.014\\
$e_9$&4.450&$e_{23}$&0.369&$e_{37}$&0.042\\
$e_{10}$&3.045&$e_{24}$&0.282&$e_{38}$&0.021\\
$e_{11}$&2.917&$e_{25}$&0.199&$e_{39}$&0.026\\
$e_{12}$&2.041&$e_{26}$&0.190&$e_{40}$&0.031\\
$e_{13}$&1.862&$e_{27}$&0.136&$e_{41}$&0.028\\
$e_{14}$&1.589&$e_{28}$&0.120&&\\
\hline
\end{tabular}
\end{center}
\end{table}

\begin{figure*}
\begin{center}
\epsfysize=18cm
\centerline {\epsfbox[32 21 564 746]{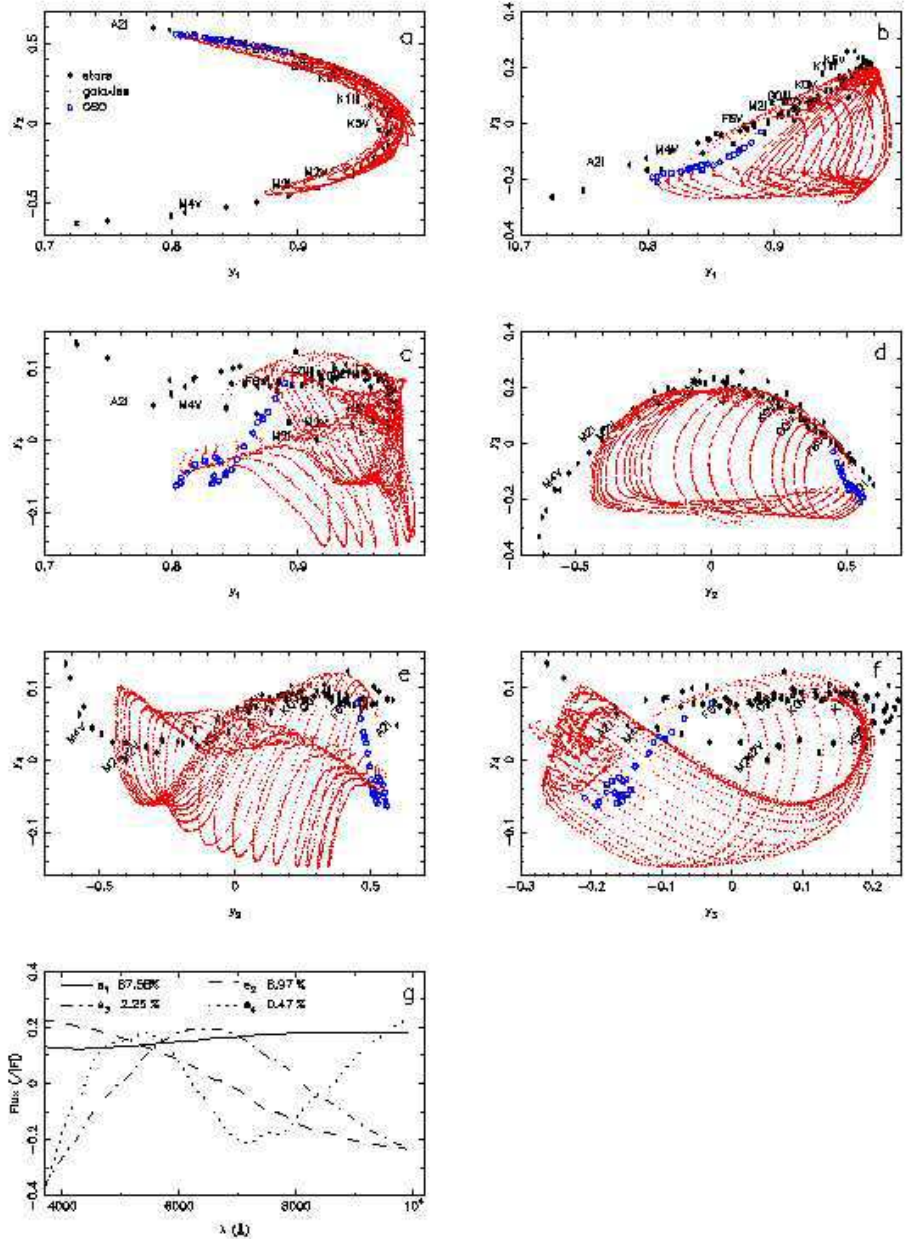}}
\caption{PCA eigencomponents $y_{2}$ versus $y_{1}$ {\bf a)}, $y_{3}$ versus $y_{1}$ {\bf b)},
$y_{4}$ versus $y_{1}$ {\bf c)}, $y_{3}$ versus $y_{2}$ {\bf d)}, $y_{4}$ versus $y_{2}$ {\bf e)},
$y_{4}$ versus $y_{3}$ {\bf f)},  and a plot of the first 4 eigenvectors
{\bf g)} for a catalog containing 131 stars ($\ast$), 7800 galaxies ($\cdot$)
and 80 QSOs ($\circ$) at a median signal-to-noise ratio of $S/N=100$ 
(see Sect{.} \protect \ref{simul}). The 3 classes of objects occupy
different regions of the graphs. Galaxies and QSOs show clear redshift sequences.
The loci of the various stellar types are indicated in each graph.}
\label{plotpcamethod}
\end{center}
\end{figure*}

Our case is more complex because we have to analyze a catalog
containing different classes of objects, which in addition span
different redshift intervals, hence showing a wide variety of spectral
energy distributions. We therefore need to use all significant
information. Several tests were performed, with PCAs having 4, 5, 6, 7, 8, 9, 10 and 20 
eigencomponents.  Fewer than 5 components is largely insufficient to describe the LZT 
mock samples (see Sect{.} 6.1). One must reach 9 components to restore
the full details of the LZT mock samples. We therefore choose to perform the
analysis of the LZT mock catalogues with a 10-eigencomponent PCA, as the remaining 
power after the 10th component just reaches below 0.1\% (see Table \ref{tablemethod} and 
Eq{.} \ref{eqgamma}). With 10 eigencomponents, Eq{.} \ref{eqM} then becomes
\begin{equation}
  f_{\lambda i}\simeq\sum_{j=1}^{10} y_{ij}e_{\lambda j}.\label{fsimp}
\end{equation}

Figure \ref{plotpcamethod} plots the first 4 weights or PCA
eigencomponents $y_{1}$, $y_{2}$, $y_{3}$, and $y_{4}$ resulting from a PCA on the 
same mock catalog that used to obtain Table \ref{tablemethod}.
The last graph shows the corresponding first 4 eigenvectors; the
corresponding relative powers which they carry are indicated in the graph.
Stars (stars) galaxies (dotted lines) and QSOs (open circles)
occupy well-defined regions of the 4-D space described by the
first 4 eigenvectors $e_1$, $e_2$, $e_3$, and $e_4$ (see frames b \& d).
As shown by \citet{connolly95b} and \citet{galaz98},
using the first 3 eigencomponents, the stellar spectral sequence (shown with asterisks) 
describes an arc, with later type being at lower values of $y_2$. 
The stellar spectral types of selected stars are indicated in each graph. 
O, B, and A stars have positive values of $y_2$, whereas M stars
have negative values of $y_2$. The other spectral types fill the sequence.
The galaxies (plotted as dots) describe a 2-D surface,
starting from the stellar sequence at low redshift, and extending away from it along
the paths generated by a redshift step of $0.01$. The higher redshift considered for the
galaxies is $\sim2$. The QSOs (open circles) describe a linear sequence defined by their
redshift ($1\le z\le4$). The QSO sequence
is distinct from the stellar sequence at low redshift, and merges with it at $z\sim3$
(see Sect{.} 4.2); at this redshift, a QSO cannot be distinguished from an F star.

Therefore, the PCA not only provides a spectral classification within
each class of object (stars, galaxies, QSOs) but also allows
a classification scheme by which the 3 classes of objects can be
distinguished. The definition of the different spectral types
for each class of object can be done by parameterization of the 1-D or
2-D surfaces they describe. This is developed in Sect{.} 3.3.

\subsection{The spectral classification} \label{class}

Because the PCA is a non-parametric approach, 
the eigenvectors derived for one catalog are specific and maybe
inappropriate for another catalog.
In order to use the PCA for spectral classification, one must
relate the internal correlations outlined by the PCA with the
physical properties of the objects.
The canonical but subjective method to achieve this purpose, is to create
realistic mock catalogues of templates and apply the PCA to them. Previous
works by \citet{connolly95a,connolly99}, \citet{galaz98}
and \citet{ronen99} establish the efficiency of the PCA to produce
an internal classification scheme related to physical properties of galaxies.
For instance, Galaz \& de Lapparent (1998) use the first 3
eigencomponents $y_{i1}$, $y_{i2}$, and $y_{i3}$ as Cartesian coordinates of a 3-D space,
and convert them into the 2 angles defining the corresponding spherical coordinates
which are in turn used for the classification: 
one parameter is related to the color of the objects, i.e. the slope of the
continua, and the other is an index of the intensity of the emission lines.
\citet{ronen99} show by using stellar synthesis galaxy template
spectra that one can track down some properties such as the age or the metallicity.
The weakness of the method of Ronen et al{.} is that when the stellar
synthesis models reproduce the observed spectra but are physically wrong, the
derived physical properties are erroneous. A successful approach 
should include a direct \emph{calibration} of the PCA on observed spectra with
known types, ages, and metallicities \citep{galaz98}.

As pointed out by \citet{connolly95b}, the PCA eigenvectors and corresponding
eigencomponents are determined by the relative numbers of the different types of objects. 
\citet{connolly95b} introduce a \emph{morphological type normalization}
to account for this relative proportion of objects when catalogues with unrealistic proportions
are used. Here, we prefer the direct approach of using
realistic mock catalogues (see Sect{.} \ref{simul}) which will be directly comparable
to the LZT catalog.

The PCA applied to the mock LZT catalogues is used here to extract the
relevant information for classifying objects and measuring their
redshifts (for galaxies and QSOs). We choose a simple
approach. Once the principal eigencomponents of the mock catalog are
measured, we select the first 10 eigenvectors and define a
10-dimensional eigenbasis, in which each class of object occupies a
given locus, defined by it spectral type among its class, and its
redshift for galaxies and QSOs. As the paths followed by the galaxies
and QSO are monotonic functions of redshift, the PCA provides a
template-independent method for measuring redshifts.  Except for the
redshift/spectral-type degeneracies discussed in Sect{.} 6.1, similar
types of objects at similar redshifts (for galaxies and QSOs) tend to
be nearby in the space defined by the 10-dimensional eigenbasis, thus
providing a unique parameterization of a given SED. In the following, we describe how the
object classes and types are defined and how the redshifts are
measured.

\subsection{Definition and measurement of object/spectral types}\label{meastype}

We define object classes (star, galaxy, QSO) and types (O to M for stars, 
E to Irr for galaxies) by assigning to each template a number $T$. 
For stars, the 131 Pickles templates spanning the range of stellar types 
O to M and luminosities I to V are ordered from blue to red and
are assigned types $T\in\{1,2,...,131\}$.
Galaxies are also ordered from blue to red as follows:
the 12 GISSEL late-type templates (see Sect{.} 2.3.2); 
the 27 PEGASE late-type templates; the 12 GISSEL early-type templates; 
the 27 PEGASE early-type templates; these
78 spectra are assigned types $T\in\{132,133,...,209\}$.
QSOs are all assigned $T=210$, as only one type of spectrum was used.

Given the SED $o_{\lambda}$ of an object. The projected
components on the eigenbasis $o^p_j$ are simple scalar products:
\begin{equation}o^p_j = o_{\lambda}\cdot e_{\lambda j}=\sum^{41}_{\lambda=1}
o_{\lambda}e_{\lambda j},\end{equation}
and the weighted {\emph{distance}} $d$ from the object to any other object $i$ 
of the 10-D space is 
\begin{equation}d=\left[\sum^{10}_{j=1}\gamma_j^{1/2}(o^p_j-y_{ij})^2\right]^{1/2}.
\label{eqchi2}\end{equation}
where $y_{ij}$ are the eigencomponents defined in Eq{.} \ref{eqM} and \ref{fsimp}, $e_j$ 
are the corresponding eigenvectors of the PCA performed on the mock catalog.
Weighting the eigencomponents with their eigenvalues $\gamma_j$ is used to account for the
increase in the noise from $y_{i1}$ to $y_{i10}$.
The object type $T_\mathrm{object}$ is simply defined as the arithmetic mean of the types
$T_{y_i}$ of the $\eta$ nearest neighbors:
\begin{equation} T_\mathrm{object} = \overline{T} = \frac{1}{\eta}\sum^{\eta}_{i=1}T_{y_i}.
\label{eqtype}\end{equation} 

As a test, we calculate for our mock catalogues of 2993 galaxies
with median $S/N=20$ and $10$, and for those with 2095 galaxies with median $S/N=6$,
the type $T_\mathrm{object}$ given by Eq{.} \ref{eqtype} 
for $\eta=1, 2, 3, 4, 5,$ and $10$. 
The resulting errors in the object classes and types are listed in 
Table \ref{table_eta}:
Cols{.} $\pm5,\pm10, \pm15$ and $>15$, give the number of galaxies for which
$T_\mathrm{object}$ differs from the true type $T$ by 5, 10, 15, and more
than 15 type units; Cols{.} star and QSO give the number of galaxies
misclassified as star, or QSO.
Table \ref{table_eta} demonstrates that in all cases, the most accurate results 
are obtained for $\eta=1$, i.e. the nearest neighbor is always the best 
choice.
The efficiency of the classification is further described in Sect{.} \ref{result}.

Note that in the nearest neighbor analysis used here, the effect
of the input $S/N$ onto the eigencomponents $y_i$ is not accounted
for. Table \ref{table_eta} suggests that the effect on the
classification would be negligeable (as almost no errors occur even
for large values of $\eta$). The effect on the type and redshift
determination would be larger, and we refer the reader to Sect{.}
\ref{degeneracy} for a more general discussion.

\begin{table}
\begin{center}
\caption{Error in the determination of the types for 2993 galaxies
at median $S/N=20$, 10, and 2095 galaxies at median $S/N=6$,
for different values of $\eta$ (see text for details). }
\label{table_eta}
\begin{tabular}{lllllll}
\multicolumn{7}{c}{Median $S/N = 20$}\\
\hline
\hline
$\eta$ & \multicolumn{4}{c}{type error \#} & \multicolumn{2}{c}{class error \#}\\
& $\pm 5$ & $\pm 10$ & $\pm 15$ & $>15$ & star& QSO\\
\hline
1 & 2796  & 14  & 20  & 133 & 0 & 0\\
2 & 2769  & 23  & 22  & 149 & 0 & 0\\
3 & 2756  & 24  & 33  & 150 & 0 & 0\\
4 & 2746  & 38  & 22  & 157 & 0 & 0\\
5 & 2726  & 56  & 27  & 154 & 0 & 0\\
10& 2703  & 50  & 50  & 160 & 0 & 0\\
\hline\\
\multicolumn{7}{c}{Median $S/N = 10$}\\
\hline
$\eta$ & \multicolumn{4}{c}{type error \#} & \multicolumn{2}{c}{class error \#}\\
& $\pm 5$ & $\pm 10$ & $\pm 15$ & $>15$ & star& QSO\\
\hline
1 & 2509  & 76  & 32  & 374 & 0 & 0\\
2 & 2469  & 75  & 43  & 404 & 0 & 0\\
3 & 2455  & 86  & 71  & 379 & 0 & 0\\
4 & 2433  & 125 & 40  & 393 & 0 & 0\\
5 & 2418  & 130 & 65  & 378 & 0 & 0\\
10& 2400  & 128 & 76  & 387 & 0 & 0\\
\hline\\
\multicolumn{7}{c}{Median $S/N = 6$}\\
\hline
$\eta$ & \multicolumn{4}{c}{type error \#} & \multicolumn{2}{c}{class error \#}\\
& $\pm 5$ & $\pm 10$ & $\pm 15$ & $>15$ & star& QSO\\
\hline
1 & 1594  & 84  & 85  & 332 & 0 & 0\\
2 & 1577  & 114 & 73  & 331 & 0 & 0\\
3 & 1587  & 105 & 88  & 315 & 0 & 0\\
4 & 1578  & 136 & 61  & 320 & 0 & 0\\
5 & 1590  & 126 & 62  & 317 & 0 & 0\\
10& 1584  & 143 & 52  & 316 & 0 & 0\\
\hline
\end{tabular}
\end{center}
\end{table}

\subsection{Measurement of redshifts}\label{measredshift}

Because we apply the PCA to the observed SEDs, it is possible to
measure redshifts directly with a technique similar to the classification
method (Sect{.} \ref{class}). In the 10-D space defined by the first 10
eigenvectors, galaxies and quasars follow redshift sequences. We define
the redshift $z_o$ of an unknown object as the average redshift $\bar{z}$
of the $\eta$ nearest neighbors from the mock catalog which
lie inside a radius $r_{\eta}$ and have  type $T_\mathrm{object}\pm 5$:
\begin{equation}z_o = \bar{z} = \frac{1}{\eta}\sum^{\eta}_{i=1}
z^{T_\mathrm{object}}_{y_i}; \label{zmeas} \end{equation}
$r_{\eta}$ is defined by $\eta$ and has the same unit as $d$ 
(Eq{.} \ref{eqchi2}), and therefore, as the eigencomponents $y_i$ (see
Eq{.} \ref{eqM} and Fig{.} \ref{plotpcamethod}). The value of $r_{\eta}$ 
is usually small: for $\eta=10$, we estimate $r_{\eta}\la0.05$ for the 3 classes
of object. 
If more than one class is present among 
the $\eta$ neighbors, i.e. if the object falls in a locus 
containing a mix of various classes, the class is not robustly 
defined and the average redshift is given for all classes inside 
$r_{\eta}$. Our mock catalog is such that for 98\% of the 
galaxies, the redshift accuracy is not strongly dependent 
on $\eta$ for $\eta<5$, and for 90\% of the QSOs,  $\eta = 1$, i.e., 
the nearest neighbor always provides a more accurate redshift 
than the average over a group of objects (the remaining 2\% galaxies 
and 10\% QSOs are affected by degeneracies
in redshifts, which are described in Sect{.} 6.1).

\section{Star/galaxy/QSO separation}\label{result}

In this section we examine quantitatively how the separation of
stars, galaxies and QSOs can be performed using the PCA, and how the
object profile can contribute to the analysis.

\subsection{Extended versus unresolved objects}\label{extended}

\begin{figure}
\begin{center}
\epsfysize=9cm
\centerline {\epsfbox[0 150 550 600]{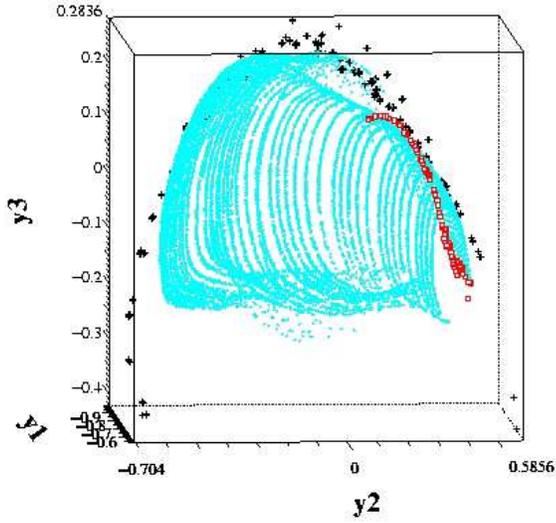}}
\caption{PCA projections $y_1$, $y_2$, and $y_3$ for a mock LZT catalog of
33,486 objects (3370 stars [+], 29,986 galaxies [$\cdot$] and 130 QSOs [squares]).
The SEDs have a median $S/N\simeq100$.}
\label{pca100}
\end{center}
\end{figure}

\begin{figure}
\begin{center}
\epsfysize=9cm
\centerline {\epsfbox[0 150 550 600]{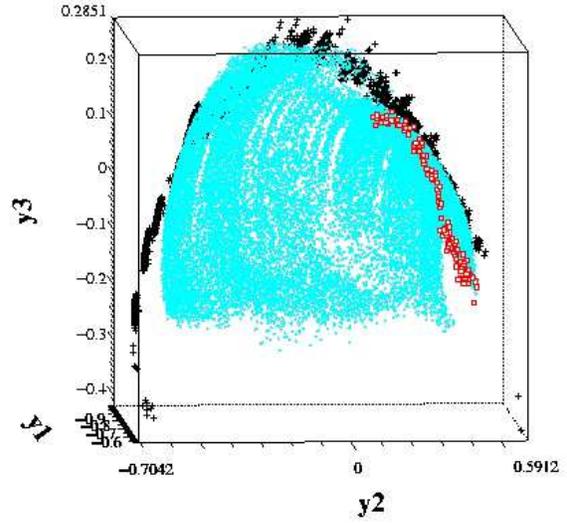}}
\caption{Same as in Fig{.} \protect\ref{pca100} for a median $S/N\simeq20$.}
\label{pca10}
\end{center}
\end{figure}

In an Einstein-de Sitter universe, the angular size $\theta$ of an object with
physical size $D$ is
\begin{equation} \theta(z) = (DH_0/2c)(1+z)^2[(1+z)-(1+z)^{1/2}]^{-1}
\end{equation}
where $c = 300,000$ km\ s$^{-1}$, and $H_0$ is the Hubble constant.
Following Sandage, Kron \& Longair \citet{sandage95},  a \emph{typical} 
disk galaxy has an effective diameter of
$D=10$ kpc, which results in an angular size $\theta(z=0.25) = 4.09
\arcsec$; this value assumes a mean surface brightness in Johnson $V$
band $\mu_V=23.5$ mag$\cdot$arcsec$^{-2}$, and an exponential disk
profile (see Table 6.2, p. 283, in Sandage, Kron \& Longair, 1995).
For $\mu_V= 25.1$ mag$\cdot$arcsec$^{-2}$, the angular size $\theta(z=0.5) = 2.8
\arcsec$.  Hence, most of bright galaxies at $z\la0.5$ are expected to
be more extended than the PSF of the LZT ($\sim1-2 \arcsec$), and
potential confusion between star and galaxy SEDs is not expected 
to occur beyond redshift $0.2$. At redshifts higher than $0.2$, the PCA is indeed
robust to segregate stars from galaxies due to the red-shifting of the
galaxy SEDs.  Only compact galaxies at low redshifts ($z\la 0.2$)  
might be intrinsically difficult to distinguish from stars in
the medium-band system of the LZT. So far there is no evidence for a
dominating population of compact galaxies at these
redshifts. Moreover, the fraction of galaxies in the predicted LZT
redshift distribution of Fig{.} \ref{totnz} lying at $z\le 0.2$ is
$\sim10$\%, a non-negligeable but minor component of the full sample.

\subsection{PCA classification efficiency}\label{stargalqsoclass}

\begin{table}
\begin{center}
\caption{Error in the classification of $\sim$ 3000 stars, $\sim$ 3000 
galaxies, and $\sim$ 1000 QSOs for different values of median signal-to-noise 
ratio $S/N$ (see text for details).}
\label{tablestargal}. 
\begin{tabular}{clllll}
\hline
\hline
Median $S/N$ & \multicolumn{2}{c}{Input $^a$}  & \multicolumn{3}{c}{Classification \# $^b$}\\
&class&\#& Stars  & Galaxies  & QSOs  \\
\hline
    & Stars    & 3181 & 3181  &  0     & 0\\
100 & Galaxies & 2993 & 0     &  2993  & 0\\
    & QSOs     & 984  & 0     &  0     & 984\\
\hline
    & Stars    & 2843 & 2833  & 10     & 0\\
20  & Galaxies & 2993 & 0     & 2993   & 0\\
    & QSOs     & 984  & 0     & 0      & 984\\
\hline
    & Stars    & 2092 & 2092  & 0      & 0\\
10  & Galaxies & 2993 & 0     & 2993   & 0\\
    & QSOs     & 984  & 6     & 0      & 978\\
\hline
    & Stars    & 1956 & 1923  & 0      & 33\\
6   & Galaxies & 2095 & 0     & 2095   & 0\\
    & QSOs     & 741  & 42    & 9     & 690\\
\hline
\end{tabular}
\end{center}
\footnotesize
$^a$ Input number of objects in each class\\
$^b$ Output number of stars, galaxies, and QSOs derived by the 10-component PCA
\end{table}

One key parameter in determining the efficiency of the PCA
classification is the signal-to-noise ratio in the spectra.  Figure
\ref{pca100} shows a 3D plot of the first 3
eigencomponents of the PCA performed on mock catalogues containing 33,486
objects (3370 stars, 29,986 galaxies and 130 QSOs) and a
median signal-to-noise ratio $S/N\simeq100$ (see Sect{.} \ref{noise}).
This catalog represents a sub-area of 1 deg$^2$ of the future LZT survey, 
and the PCA method is that described in Sect{.} \ref{class}, based on 
the first 10 eigencomponents. In Fig{.} \ref{pca10}, the mock
catalog of Fig{.} \ref{pca100} is degraded to a median signal-to-noise ratio
$S/N\simeq20$. In both Figs{.}, stars, galaxies and QSOs occupy different loci 
in the 3D-space, therefore allowing us to separate objectively the 3 classes 
of objects, and illustrating how the PCA is robust to noise.

Table \ref{tablestargal} compiles the efficiency of the classification as a 
function of signal-to-noise ratio for a small mock catalog described in 
Sect{.} 5 below.
Following the previous Sect{.}, we assume that the 1347
galaxies at $z\la0.5$ are extended, and thus can be
separated from stars using morphological criteria. 
The remaining 1646 galaxies at $z\ga0.5$
are separated from stars and QSOs using Eq{.} \ref{eqtype}.
For a median $S/N=100$, the object classification is perfect, and 
although it slowly degrades as the signal-to-noise ratio decreases,
the performance of the classification are weakly dependent on
the signal-to-noise ratio.

Table \ref{tablestargal} shows that QSO SEDs tend to be misclassified as 
stellar SEDs ($6\%$ of QSOs at a median signal-to-noise ratio of 6). 
There is no degeneracy between blue stars (O, B or white dwarfs) and 
low-redshift QSOs, because blue stars have no emission lines; QSOs are thus  
clearly segregated. In the range $2.5<z<3$, QSO and F/G stellar SEDs 
are similar.
For $z\ga 3.5$, QSO signatures are unique again and misclassification is rare.
Unfortunately, the range $2.5<z<3$ corresponds to the peak of the expected QSO
distribution (see Sect{.} \ref{QSOs}). This explains why such a large fraction of 
QSOs are misclassified as stars for median $S/N=6$.
This outlines a weakness of this classification method, the solution of which
probably lies in a slightly different approach which we
discuss in Sect{.} \ref{discussion}.

\begin{table}
\begin{center}
\caption{Error in the determination of the type for stars and galaxies,
for different values of median signal-to-noise ratio $S/N$.
Stars have a total of 131 types, and galaxies have 78 types (see text for details). }
\label{tableerrtype}
\begin{tabular}{cllllll}
\hline
\hline
Median $S/N$ & Object class & \multicolumn{4}{c}{type error \# $^a$} \\
    &              & $\pm 5$ & $\pm 10$ & $\pm 15$ & $>15$\\
\hline
100 & Stars    & 3177 & 3   & 0  & 1\\
    & Galaxies & 2916 & 0   & 4  & 25\\
\hline
20  & Stars    & 2617 & 163 & 33 & 30\\
    & Galaxies & 2796 & 14  & 20 & 133\\
\hline
10  & Stars    & 1658 & 290 & 73 & 71\\
    & Galaxies & 2509 & 76  & 32 & 374\\
\hline
6   & Stars    & 1560 & 317 & 38 & 41\\
    & Galaxies & 1594 & 85  & 84 & 332\\
\hline
\end{tabular}
\end{center}
\footnotesize
$^a$ Number of objects classified within $\pm 5$, $\pm 10$, $\pm 15$ and more than
$\pm 15$ of their input types
\end{table}

\subsection{PCA type identifications}\label{typemeasure}

Table \ref{tableerrtype} allows us to evaluate the accuracies in the type 
identifications for the 2 classes of
objects having types (stars and galaxies).
The accuracies are measured by the difference between the real type
and the PCA-measured type, as in Table \ref{table_eta}. 
For galaxies, an error of $\pm 10$ on the type leads to a catastrophic 
error in redshift in $1/3$ of the cases, and an error of $\pm 15$ or more
leads systematically to a catastrophic error in redshift. 
Table \ref{tableerrtype} shows the results for the 4 
median signal-to-noise ratios $S/N=100, 20, 10, 6$.\\
At a median $S/N=100$, the PCA is able to discriminate O,B,A,F,G,K,M star types
and most luminosity and metallicity classes. 
At a median $S/N=20$, metallicity differences within each stellar type 
are no longer discriminated, and the luminosity class I, II, III, IV, V 
is often mismatched. At a median $S/N=10$ and lower, only major continuum differences 
allow one to discriminate between different stellar types. 
Table \ref{tableerrtype} shows that even at $S/N=6$, the PCA is able to discriminate
types for half of the objects within each class.

\section{Redshift measurement accuracy}\label{redshift}

\begin{figure*}
\begin{center}
\epsfysize=15cm
\centerline {\epsfbox[35 23 552 737]{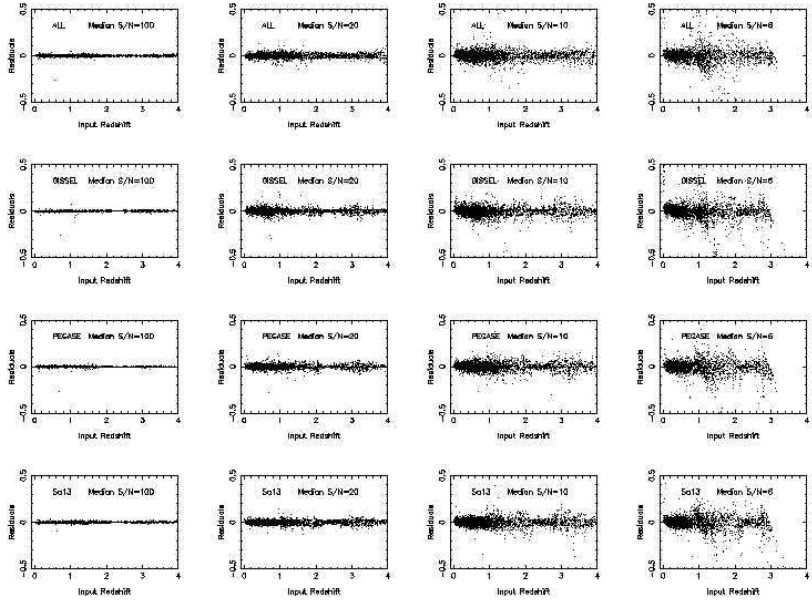}}
\caption{Redshift residuals versus input redshift extracted from a $\chi^2$
reduction of the first 10 PCA eigencomponents. In each frame, spectra in 
sub-mock catalogues are projected onto realistic mock simulations with a 
median signal-to-noise ratio of $S/N=100$.
>From top to bottom, in addition to $\sim3000$ stars and $\sim1000$ QSOs, 
the sub-mock calatogues include $\sim3000$ galaxy SEDs drawn respectively from 
only the 78 galaxy templates of PEGASE+GISSEL (ALL), 
from only the 24 galaxy templates of GISSEL, from only the 54 PEGASE templates (PEGASE), and from only
the 31 PEGASE Sa13 templates (Sa13). In all frames, only QSOs have redshifts $z>2$.
The corresponding standard deviations are
listed in Tables \protect\ref{redshifts100} to \protect\ref{redshifts6}.}
\label{plotz}
\end{center}
\end{figure*}

\begin{table*}
\begin{center}
\caption{Residuals in the redshift measurements for sub-mock simulations including $\sim 3000$ stars, $\sim 1000$ QSOs, and $\sim 3000$ galaxy SEDs with a median signal-to-noise ratio 
of 100 (see also Fig{.} \protect\ref{plotz}).}
\label{redshifts100}
\begin{tabular}{lllllllllllll}
\hline
\hline
\multicolumn{13}{c}{Median $S/N=100$}\\
\hline
$z$&\multicolumn{3}{c}{ALL}&\multicolumn{3}{c}{GISSEL}&\multicolumn{3}{c}{PEGASE}&\multicolumn{3}{c}{PEGASE Sa13}\\
&\#&$\bar{z}$&$\sigma_\mathrm{Res.}$&\#&$\bar{z}$&$\sigma_\mathrm{Res.}$&\#&$\bar{z}$&$\sigma_\mathrm{Res.}$&\#&$\bar{z}$&$\sigma_\mathrm{Res.}$\\
\hline
0.1 & 243 & -0.0002 & 0.006  & 244 & -0.0003 & 0.004  & 243 & -0.0002 & 0.004 & 251 & 0.0004 & 0.004\\
0.2 & 334 & -0.0001 & 0.004  & 336 & -0.0001 & 0.003 & 325 & 0.0001 & 0.003 & 329 & 0.0000 & 0.003\\
0.3 & 354 & -0.0004 & 0.014  & 358 & -0.0003 & 0.004 & 358 & -0.0003 & 0.004 & 359 & -0.0003 & 0.003\\
0.4 & 357 & 0.0001 & 0.005 & 354 & -0.0004 & 0.014 & 326 & -0.0004 & 0.004 & 351 & 0.0003 & 0.005\\
0.5 & 319 & -0.0001 & 0.005 & 318 & -0.0005 & 0.004 & 288 & -0.0001 & 0.004 & 325 & 0.0002 & 0.006\\
0.6 & 293 & -0.0005 & 0.004& 291 & -0.0001 & 0.004  & 258 & 0.0001 & 0.004 & 288 & 0.0001 & 0.004\\
0.7 & 254 & 0.0004 & 0.004  & 251 & -0.0003 & 0.003  & 221 & 0.0004 & 0.005 & 254 & -0.0001 & 0.004\\
0.8 & 219 & -0.0006 & 0.005  & 225 & 0.0001 & 0.003  & 177 & -0.0001 & 0.006 & 224 & 0.0001 & 0.004 \\
0.9 & 181 & 0.0002 & 0.005  & 181 & 0.0003 & 0.004 & 147 & 0.0008 & 0.005  & 178 & -0.0002 & 0.006\\
1.0 & 148 & 0.0002 & 0.008  & 146 & 0.0010 & 0.010 & 114 & -0.0012 & 0.005  & 150 & -0.0005 & 0.006\\
1.1 & 111 & -0.0004 & 0.007 & 114 & -0.0005 & 0.005 & 90 & -0.0003 & 0.006 & 110 & 0.0007 & 0.006\\
1.2 & 90 & -0.0002 & 0.007  & 88 & -0.0010 & 0.004  & 66 & -0.0005 & 0.009  & 90 & 0.0003 & 0.006\\
1.3 & 68 & -0.0009 & 0.010 & 68 & -0.0007 & 0.004  & 52 & -0.0012 & 0.008  & 64 & 0.0003 & 0.008 \\
1.4 & 51 & 0.0006 & 0.009  & 49 & -0.0009 & 0.004  & 47 & -0.0001 & 0.009 & 54 & -0.0045 & 0.006  \\
1.5 & 48 & -0.0010 & 0.011  & 50 & 0.0001 & 0.003  & 41 & 0.0000 & 0.008  & 47 & 0.0009 & 0.007\\
1.6 & 40 & 0.0009 & 0.010  & 40 & 0.0003 & 0.005   & 39 & -0.0025 & 0.009  & 40 & 0.0002 & 0.008\\
1.7 & 39 & 0.0027 & 0.007  & 39 & -0.0011 & 0.006   & 35 & -0.0026 & 0.007  & 39 & 0.0004 & 0.007\\
1.8 & 35 & -0.0003 & 0.008 & 35 & -0.0003 & 0.007  & 31 & -0.0010 & 0.006  & 35 & -0.0031 & 0.007\\
1.9 & 34 & -0.0000 & 0.007  & 29 & -0.0017 & 0.005  & 31 & -0.0010 & 0.006& 32 & 0.0003 & 0.007\\
2.0 & 40 & -0.0005 & 0.006  & 45 & 0.0024 & 0.007  & 43 & 0.0005 & 0.007  & 42 & -0.0010 & 0.007\\
2.5 & 44 & 0.0005 & 0.004  & 48 & -0.0002 & 0.006   & 50 & -0.0008 & 0.006 & 46 & 0.0009 & 0.005\\
3.0 & 40 & -0.0002 & 0.006  & 39 & 0.0021 & 0.007 & 39 & -0.0031 & 0.007  & 42 & -0.0005 & 0.005\\
3.5 & 31 & -0.0010 & 0.005  & 27 & -0.0004 & 0.005  & 28 & 0.0025 & 0.008 & 32 & 0.0025 & 0.006 \\

\hline
\end{tabular}
\end{center}
\footnotesize
\underline{Notes:} \\
at $z<2$, $z$ is the redshift of the galaxies binned by intervals of 0.1\\
at $z\ge2$, $z$ is the redshift of the QSOs binned by intervals of 0.5\\
\# is the number of galaxies or QSOs in the redshift bin considered\\
$\bar{z}$ is the average residual in the redshift\\
$\sigma_\mathrm{Res.}$ is the r{.}m{.}s{.} dispersion in the redshifts errors defined as $\Delta z=|z_\mathrm{PCA}-z_\mathrm{input}|$\\
ALL corresponds to sub-mock catalogues with galaxies drawn from all galaxy templates\\
GISSEL corresponds to sub-mock catalogues with galaxies drawn from GISSEL templates only\\
PEGASE corresponds to sub-mock catalogues with galaxies drawn from PEGASE templates only\\ 
PEGASE Sa13 corresponds to sub-mock catalogues with galaxies drawn from PEGASE Sa13 templates only
\end{table*}

\begin{table*}
\begin{center}
\caption{Same as in Table \protect\ref{redshifts100} for a median signal-to-noise
ratio of 20 (see also Fig{.} \protect\ref{plotz}).}
\label{redshifts20}
\begin{tabular}{lllllllllllll}
\hline
\hline
\multicolumn{13}{c}{Median $S/N=20$}\\
\hline
$z$&\multicolumn{3}{c}{ALL}&\multicolumn{3}{c}{GISSEL}&\multicolumn{3}{c}{PEGASE}&\multicolumn{3}{c}{PEGASE Sa13}\\
&\#&$\bar{z}$&$\sigma_{Res.}$&\#&$\bar{z}$&$\sigma_{Res.}$&\#&$\bar{z}$&$\sigma_{Res.}$&\#&$\bar{z}$&$\sigma_{Res.}$\\
\hline
0.1 & 242 & -0.0012 & 0.014 & 242 & -0.0002 & 0.016 & 245 & 0.0003 & 0.015 & 248 & -0.0008 & 0.012 \\
0.2 & 330 & -0.0010 & 0.013  & 335 & 0.0008 & 0.015 & 333 & -0.0010 & 0.012 & 330 & 0.0005 & 0.012\\
0.3 & 360 & -0.0002 & 0.013  & 354 & -0.0003 & 0.015 & 351 & -0.0000 & 0.013 & 357 & -0.0009 & 0.013\\
0.4 & 353 & 0.0009 & 0.016 & 355 & -0.0027 & 0.030 & 357 & -0.0017 & 0.021 & 354 & 0.0010 & 0.014\\
0.5 & 319 & 0.0023 & 0.019  & 317 & -0.0019 & 0.024 & 322 & -0.0009 & 0.018 & 320 & 0.0010 & 0.016\\
0.6 & 295 & 0.0015 & 0.017  & 296 & 0.0007 & 0.028 & 293 & -0.0002 & 0.016 & 290 & -0.0004 & 0.014\\
0.7 & 251 & 0.0004 & 0.019  & 251 & -0.0015 & 0.021 & 253 & -0.0001 & 0.018 & 257 & -0.0005 & 0.013\\
0.8 & 219 & -0.0021 & 0.021  & 220 & -0.0013 & 0.021 & 217 & 0.0001 & 0.019 & 218 & 0.0003 & 0.015\\
0.9 & 181 & 0.0020 & 0.020  & 179 & 0.0027 & 0.021 & 183 & -0.0020 & 0.019 & 180 & -0.0009 & 0.018 \\
1.0 & 147 & 0.0023 & 0.024 & 155 & -0.0004 & 0.021 & 149 & 0.0031 & 0.026 & 149 & 0.0008 & 0.026\\
1.1 & 116 & 0.0034 & 0.026 & 108 & 0.0064 & 0.039 & 112 & -0.0016 & 0.022 & 114 & -0.0008 & 0.025\\
1.2 & 86 & -0.0008 & 0.025  & 91 & 0.0007 & 0.020 & 87 & 0.0001 & 0.026 & 88 & -0.0035 & 0.026\\
1.3 & 66 & 0.0005 & 0.032  & 66 & -0.0014 & 0.015 & 67 & 0.0032 & 0.021 & 66 & 0.0091 & 0.023\\
1.4 & 53 & -0.0023 & 0.025  & 50 & -0.0023 & 0.015 & 51 & 0.0046 & 0.018 & 52 & -0.0027 & 0.019\\
1.5 & 48 & 0.0057 & 0.082 & 49 & 0.0006 & 0.015 & 48 & -0.0019 & 0.025 & 49 & -0.0027 & 0.032\\
1.6 & 40 & -0.0052 & 0.032  & 39 & 0.0004 & 0.019 & 40 & 0.0003 & 0.023 & 40 & -0.0067 & 0.020\\
1.7 & 39 & 0.0171 & 0.114  & 40 & 0.0053 & 0.041  & 40 & 0.0091 & 0.029 & 39 & -0.0047 & 0.027\\
1.8 & 35 & -0.0051 & 0.025  & 35 & 0.0017 & 0.032 & 35 & -0.0049 & 0.027 & 35 & -0.0034 & 0.031\\
1.9 & 33 & -0.0021 & 0.023  & 34 & 0.0047 & 0.027 & 32 & -0.0100 & 0.033 & 31 & -0.0094 & 0.026\\
2.0 & 41 & 0.0034 & 0.022  & 40 & 0.0027 & 0.026 & 42 & -0.0036 & 0.029 & 43 & 0.0023 & 0.030 \\
2.5 & 46 & -0.0030 & 0.013  & 44 & 0.0005 & 0.013 & 46 & 0.0011 & 0.016 & 47 & 0.0028 & 0.012\\
3.0 & 38 & -0.0037 & 0.019  & 39 & -0.0021 & 0.034 & 37 & 0.0016 & 0.029 & 42 & 0.0012 & 0.033\\
3.5 & 28 & -0.0068 & 0.017  & 29 & -0.0117 & 0.023 & 29 & 0.0100 & 0.020 & 29 & 0.0041 & 0.018\\
\hline
\\
\end{tabular}
\end{center}

\begin{center}
\caption{Same as in Table \protect\ref{redshifts100} for a median signal-to-noise
ratio of 10 (see also Fig{.} \protect\ref{plotz}).}
\label{redshifts10}
\begin{tabular}{lllllllllllll}
\hline
\hline
\multicolumn{13}{c}{Median $S/N=10$}\\
\hline
$z$&\multicolumn{3}{c}{ALL}&\multicolumn{3}{c}{GISSEL}&\multicolumn{3}{c}{PEGASE}&\multicolumn{3}{c}{PEGASE Sa13}\\
&\#&$\bar{z}$&$\sigma_{Res.}$&\#&$\bar{z}$&$\sigma_{Res.}$&\#&$\bar{z}$&$\sigma_{Res.}$&\#&$\bar{z}$&$\sigma_{Res.}$\\
\hline
0.1 & 243 & -0.0006 & 0.026  & 245 & 0.0072 & 0.030 & 241 & -0.0002 & 0.023 & 242 & -0.0010 & 0.022\\
0.2 & 333 & 0.0019 & 0.028  & 327 & 0.0007 & 0.032 & 329 & 0.0013 & 0.020 & 333 & -0.0010 & 0.023\\
0.3 & 356 & -0.0030 & 0.027  & 360 & -0.0008 & 0.034 & 360 & 0.0010 & 0.025 & 357 & 0.0015 & 0.025\\
0.4 & 358 & -0.0025 & 0.032 & 353 & -0.0089 & 0.039  & 357 & -0.0011 & 0.030 & 347 & -0.0016 & 0.034\\
0.5 & 320 & -0.0038 & 0.033 & 325 & -0.0078 & 0.039  & 317 & -0.0023 & 0.029 & 323 & 0.0006 & 0.029\\
0.6 & 290 & -0.0020 & 0.031 & 289 & -0.0070 & 0.044  & 296 & 0.0025 & 0.034 & 296 & 0.0000 & 0.025\\
0.7 & 256 & 0.0022 & 0.039  & 250 & -0.0053 & 0.036 & 250 & 0.0025 & 0.037 & 249 & -0.0010 & 0.027\\
0.8 & 218 & -0.0023 & 0.037  & 221 & -0.0096 & 0.041 & 223 & 0.0004 & 0.032& 227 & -0.0043 & 0.031\\
0.9 & 179 & -0.0016 & 0.043 & 179 & -0.0090 & 0.050  & 178 & 0.0015 & 0.036& 175 & -0.0017 & 0.032\\
1.0 & 153 & -0.0005 & 0.085  & 152 & -0.0103 & 0.062 & 146 & 0.0002 & 0.04 & 151 & 0.0052 & 0.0436\\
1.1 & 108 & -0.0041 & 0.068  & 113 & -0.0119 & 0.045 & 115 & 0.0010 & 0.047& 112 & -0.0008 & 0.050\\
1.2 & 92 & 0.0184 & 0.102  & 86 & 0.0081 & 0.044 & 89 & -0.0034 & 0.061 & 89 & -0.0038 & 0.053\\
1.3 & 67 & -0.0006 & 0.042  & 68 & -0.0008 & 0.038 & 66 & 0.0063 & 0.042 & 66 & 0.0158 & 0.056\\
1.4 & 49 & -0.0038 & 0.045  & 50 & -0.0059 & 0.035 & 49 & 0.0005 & 0.046 & 50 & 0.0002 & 0.043\\
1.5 & 50 & -0.0121 & 0.066 & 48 & 0.0030 & 0.059  & 51 & 0.0066 & 0.068 & 49 & -0.0024 & 0.084\\
1.6 & 40 & 0.0385 & 0.165 & 40 & -0.0007 & 0.107  & 39 & 0.0388 & 0.143 & 40 & 0.0028 & 0.041\\
1.7 & 39 & 0.0404 & 0.196  & 40 & 0.0199 & 0.108 & 40 & 0.0380 & 0.143 & 40 & 0.0142 & 0.105\\
1.8 & 35 & 0.0017 & 0.094 & 35 & 0.0083 & 0.044  & 35 & -0.0080 & 0.052 & 35 & -0.0046 & 0.056\\
1.9 & 27 & 0.0411 & 0.124  & 33 & 0.0112 & 0.056 & 31 & -0.0010 & 0.053 & 33 & -0.0061 & 0.078\\
2.0 & 47 & 0.0217 & 0.113  & 41 & -0.0051 & 0.058 & 43 & 0.0116 & 0.057 & 41 & -0.0012 & 0.060\\
2.5 & 44 & -0.0032 & 0.032  & 46 & 0.0083 & 0.030 & 44 & 0.0023 & 0.025 & 46 & 0.0028 & 0.023\\
3.0 & 41 & 0.0002 & 0.031  & 41 & -0.0078 & 0.074 & 39 & -0.0115 & 0.068 & 42 & -0.0155 & 0.094\\
3.5 & 28 & 0.0014 & 0.042  & 32 & 0.0059 & 0.056 & 32 & 0.0206 & 0.041 & 29 & -0.0045 & 0.048\\
\hline
\\
\end{tabular}
\end{center}
\end{table*}

\begin{table*}
\begin{center}
\caption{Same as in Table \protect\ref{redshifts100} for a median signal-to-noise
ratio of 6 (see also Fig{.} \protect\ref{plotz}).}
\label{redshifts6}
\begin{tabular}{lllllllllllll}
\hline
\hline
\multicolumn{13}{c}{Median $S/N=6$}\\
\hline
$z$&\multicolumn{3}{c}{ALL}&\multicolumn{3}{c}{GISSEL}&\multicolumn{3}{c}{PEGASE}&\multicolumn{3}{c}{PEGASE Sa13}\\
&\#&$\bar{z}$&$\sigma_{Res.}$&\#&$\bar{z}$&$\sigma_{Res.}$&\#&$\bar{z}$&$\sigma_{Res.}$&\#&$\bar{z}$&$\sigma_{Res.}$\\
\hline
0.1 & 216 & 0.0055 & 0.026 & 217 & 0.0126 & 0.031  & 217 &0.0029 & 0.030 & 208 & 0.0026 & 0.028\\
0.2 & 269 & 0.0028 & 0.033 & 262 & 0.0065 & 0.039& 261 & -0.0007 & 0.032 & 266 & 0.0010 & 0.036\\
0.3 & 260 & 0.0010 & 0.037  & 267 & 0.0070 & 0.113& 266& 0.0013 & 0.033 & 270 & -0.0002 & 0.030\\
0.4 & 222 & -0.0007 & 0.099  & 223 & 0.0013 & 0.069& 222 &-0.0029 & 0.050 & 220 & 0.0039 & 0.051 \\
0.5 & 175 & 0.0004 & 0.053  & 173 & 0.0014 & 0.067& 177 &  0.0009 & 0.064 & 177 & -0.0021 & 0.039\\
0.6 & 129 & 0.0018 & 0.065  & 131 & -0.0027 & 0.082& 128 & 0.0035 & 0.067 & 131 & 0.0048 & 0.045\\
0.7 & 90 & -0.0008 & 0.052  & 89 & 0.0012 & 0.063& 91 & 0.0043 & 0.042 & 86 & 0.0045 & 0.039\\
0.8 & 118 & -0.0807 & 0.185  & 122 & -0.0223 & 0.091& 58 & 0.0038 & 0.045 & 60 & 0.0027 & 0.047\\
0.9 & 184 & -0.1366 & 0.237  & 174 & -0.0654 & 0.146 & 102 & -0.0253 & 0.124 & 101 & -0.0113 & 0.091\\
1.0 & 144 & -0.0965 & 0.249  & 155 & -0.0812 & 0.156 & 146 & -0.0568 & 0.168 & 148 & 0.0007 & 0.092\\
1.1 & 116 & -0.0639 & 0.235  & 107 & -0.0132 & 0.089 & 112 & -0.0248 & 0.122 & 113 & -0.0036 & 0.098\\
1.2 & 87 & 0.0239 & 0.207  & 91 & -0.0076 & 0.129 & 89 &  -0.0248 & 0.122 & 89 & -0.0108 & 0.091\\
1.3 & 66 & 0.0209 & 0.244  & 63 & 0.0265 & 0.133 & 67 & -0.0043 & 0.114 & 65 & -0.0013 & 0.103\\
1.4 & 50 & 0.0617 & 0.233  & 50 & -0.0102 & 0.093 & 51 & -0.0042 & 0.148 & 52 & 0.0058 & 0.083\\
1.5 & 49 & 0.0577 & 0.266  & 52 & 0.0114 & 0.131 & 50 & 0.0117 & 0.111 & 49 & -0.0024 & 0.111\\
1.6 & 41 & 0.1385 & 0.339  & 40 & 0.0030 & 0.110 & 39 & 0.0313 & 0.151 & 39 & -0.0060 & 0.078\\
1.7 & 39 & 0.0411 & 0.248  & 39 & -0.0152 & 0.123 & 40 & 0.0186 & 0.184 & 40 & 0.0388 & 0.140\\
1.8 & 35 & 0.1317 & 0.355  & 35 & -0.0077 & 0.109 & 35 & 0.0211 & 0.081 & 35 & -0.0220 & 0.122\\
1.9 & 34 & 0.1615 & 0.354  & 31 & -0.0210 & 0.100  & 32 & 0.0062 & 0.089 & 34 & -0.0141 & 0.082\\
2.0 & 39 & 0.0182 & 0.186  & 43 & 0.0056 & 0.081  & 42 & 0.0129 & 0.084 & 40 & -0.0185 & 0.073\\
2.5 & 39 & 0.0377 & 0.154  & 47 & -0.0117 & 0.074 & 45 & -0.0276 & 0.087 & 48 & -0.0137 & 0.070\\
3.0 & 35 & 0.0111 & 0.054 & 38 & -0.0132 & 0.116 & 38 & -0.0216 & 0.097 & 38 & -0.0145 & 0.120\\
\hline
\end{tabular}
\end{center}

\begin{center}
\caption{Residuals in the galaxy redshift measurements binned by increasing redshift for
the simulations including GISSEL templates only on a PCA including
PEGASE templates only, for various median signal-to-noise ratios
(see also Fig{.} \protect\ref{zsys.ps}).}
\label{redshiftzsys}
\begin{tabular}{lllllllllllll}
\hline
\hline
$z$&\multicolumn{3}{c}{Median $S/N=100$}&\multicolumn{3}{c}{Median $S/N=20$}&\multicolumn{3}{c}{Median $S/N=10$}&\multicolumn{3}{c}{Median $S/N=6$}\\
  &\#&$\bar{z}$&$\sigma_{Res.}$&\#&$\bar{z}$&$\sigma_{Res.}$&\#&$\bar{z}$&$\sigma_{Res.}$\\
\hline
0.1 & 244 & -0.0024 & 0.021  & 242 & -0.0024 & 0.023& 245 & -0.0013 & 0.033 & 217 & 0.0066 & 0.041\\
0.2 & 336 & -0.0012 & 0.022  & 335 & -0.0031 & 0.023 & 327 & -0.0068 & 0.034 & 262 & -0.0107 & 0.056\\
0.3 & 358 & -0.0022 & 0.028  & 354 & -0.0038 & 0.033 & 360 & -0.0112 & 0.040 & 267 & -0.0160 & 0.076\\
0.4 & 354 & -0.0128 & 0.044  & 355 & -0.0125 & 0.046 & 353 & -0.0249 & 0.050 & 223 & -0.0313 & 0.059\\
0.5 & 318 & -0.0206 & 0.049  & 317 & -0.0172 & 0.049 & 325 & -0.0263 & 0.051 & 173 & -0.0392 & 0.056\\
0.6 & 247 & -0.0183 & 0.047  & 286 & -0.0252 & 0.050 & 289 & -0.0329 & 0.055 & 131 & -0.0508 & 0.062\\
0.7 & 184 & -0.0459 & 0.048  & 251 & -0.0426 & 0.046 & 251 & -0.0468 & 0.055 & 89 & -0.0557 & 0.101\\
0.8 & 157 & -0.0581 & 0.057  & 220 & -0.0557 & 0.049 & 221 & -0.0834 & 0.124 & 122 & -0.1634 & 0.220\\
0.9 & 124 & -0.1432 & 0.220  & 161 & -0.1066 & 0.176 & 179 & -0.1298 & 0.204 & 174 & -0.3560 & 0.284\\
1.0 & 86 & -0.1900 & 0.254  & 110 & -0.1503 & 0.213 & 152 & -0.1829 & 0.238 & 155 & -0.3805 & 0.292\\
1.1 & 77 & -0.1933 & 0.267  & 77 & -0.1825 & 0.261 & 113 & -0.2285 & 0.273 & 107 & -0.3045 & 0.313\\
1.2 & 59 & -0.1734 & 0.257  & 66 & -0.1248 & 0.201 & 69 & -0.1534 & 0.243 & 91 & -0.2792 & 0.303\\
1.3 & 55 & -0.0651 & 0.133  & 51 & -0.0559 & 0.119 & 51 & -0.0519 & 0.145 & 63 & -0.0640 & 0.148\\
1.4 & 42 & -0.0520 & 0.103  & 40 & -0.0501 & 0.080 & 34 & -0.0239 & 0.101 & 50 & -0.0582 & 0.163\\
1.5 & 41 & 0.0264 & 0.134  & 44 & 0.0315 & 0.152 & 41 & 0.0053 & 0.127 & 52 & 0.0000 & 0.165\\
1.6 & 36 & 0.0833 & 0.191  & 36 & 0.0123 & 0.133 & 37 & 0.0523 & 0.196 & 36 & 0.0203 & 0.167\\
1.7 & 37 & 0.0601 & 0.191  & 37 & 0.0558 & 0.189 & 37 & 0.0423 & 0.177 & 39 & 0.0809 & 0.253\\
\hline
\end{tabular}
\end{center}
\footnotesize
\underline{Definition of Cols{.}:} \\
\#: number of galaxies or QSOs in the redshift bin considered\\
$\bar{z}$: average residual in the redshift\\
$\sigma_{Res.}$: standard deviation in residual
\end{table*}

In order to test the accuracy of the measurement of redshifts with
the PCA, we must make a prior hypothesis that the mock catalog is
a fair representation of real observations. In that case, it is
acceptable to compare sub-samples of the main catalog to itself, for
different $S/N$ ratio to verify that we can recover physical information
(here, the redshift). In other words, internal errors will be
a good representation of the errors. On the other hand, if the mock catalog
is not a fair simulation of the observations, additional systematic errors
will degrade the measured accuracy. The underlying motivation for using sub-samples
of the mock catalog (denoted ``sub-mock'' catalogues hereafter) is obviously to save computing time.
Keeping this remark in mind, we generate small mock catalogues of galaxy
SEDs according to the procedure described in Sect{.} \ref{simul}.
The small mock catalogues contain $\sim3000$ stars, $\sim3000$ galaxies and
$\sim1000$ QSOs. We then project the SEDs onto the first 10 eigenspectra 
derived from PCAs made on mock catalogues with realistic proportions of stars, galaxies 
QSOs ($\sim 3000$ stars, $\sim 30\,000$ galaxies, $\sim 1000$ QSOs),
and with a median signal-to-noise ratio $S/N=100$. The resulting 10 eigencomponents
of each SED are compared to the corresponding eigencomponents of
the realistic mock catalog using the least-square technique described
in Sect{.} \ref{method} (Eqs{.} \ref{eqchi2} and \ref{zmeas}).

Figure \ref{plotz} (frames labeled ``ALL'') shows the residuals in the
redshift measurement and Tables \ref{redshifts100} to \ref{redshifts6}
show the standard deviations in the residuals of measured redshifts 
versus input redshifts, as a function of redshift using all galaxy 
templates available (54 PEGASE templates, 24 GISSEL templates).
In order to evaluate the impact on the error budget of using different
templates for the galaxy SEDs, we restrict the sub-mock catalogues, whereas
the eigenbasis on which the various sub-samples are projected remains 
unchanged and contains all 54 PEGASE templates and 24 GISSEL templates.
For each median signal-to-noise value, we generate sub-mock catalogues 
using for galaxies: only the 24 GISSEL templates
(Tables \ref{redshifts100} to \ref{redshifts6}, and frames
labeled ``GISSEL'' in Fig{.} \ref{plotz}); only the 54 PEGASE templates (Tables \ref{redshifts100}
to \ref{redshifts6}, and frames labeled ``PEGASE'' in Fig{.} \ref{plotz});
only the 31 Sa13 PEGASE templates (Tables \ref{redshifts100} to
\ref{redshifts6}, Cols{.} labeled ``PEGASE Sa13'', and frames labeled Sa13 in Fig{.} \ref{plotz}).  
In Fig{.} \ref{plotz}, the simulations are shown for median signal-to-noise 
ratios $S/N=100$, $S/N=20$, $S/N=10$, and $S/N=6$. 

All groups of templates shown in Fig{.} \ref{plotz} and Tables 
\ref{redshifts100} to \ref{redshifts6} show similar trends. They
yield accurate redshift measurements at all redshifts for median $S/N=100, 20$,
with $\sigma_\mathrm{Res.}\la 0.02$ for galaxies at $0.1\le z\le1.2$ or 
$1.8\le z<2.0$ and all QSOs (at $z\ge2.0$), and 
$\sigma_\mathrm{Res.}\simeq0.02-0.04$ for galaxies in $1.3\le z\le1.7$. 
For a median $S/N=10$, the errors in the redshift measurement 
remain very small with $\sigma_\mathrm{Res.}\la 0.04$ for galaxies at 
$0.1\le z\le0.9$, $\sigma_\mathrm{Res.}\simeq 0.03-0.1$ for QSOs (at $z\ge2.0$),
and $\sigma_\mathrm{Res.}\simeq0.03-0.2$ for galaxies in $1.0\le z<2.0$. 
Even for the lowest signal-to-noise ratio of $S/N=6$, the redshift
measurement is robust to $z\le0.7$ for galaxies, with
$\sigma_\mathrm{Res.}\simeq 0.03-0.1$, and then suffers from catastrophic
degeneracies at $z\ge0.9$, where $\sigma_\mathrm{Res.}$ grows to
$\sim0.2-0.3$; similar problems affect QSOs, which
have $\sigma_\mathrm{Res.}\simeq0.05-0.2$ for $2\le z\le 3.0$.

Note that the dispersion in the redshift residuals for median $S/N=10$ and
$S/N=6$ are comparable with the results obtained by \citet{hickson94a}
using a $\chi^2$ adjustment of mock LZT galaxy SEDs onto the PEGASE 
templates; in contrast to the results of \citet{hickson94a}, the 
dispersion in the PCA redshift residuals increases monotonically with redshift.
Tables \ref{redshifts100} to \ref{redshifts6} also show that
the intense emission lines present in the QSOs allow us to measure redshifts 
at all signal-to-noise ratios. The problem of possible misclassification 
of QSO as stars, mentioned in Sect{.} \ref{stargalqsoclass}, 
only affects the completeness of the sample, but 
does not affect the quality of redshift measurements.

The similar redshift residuals for the 4 types of sub-mock catalogues
(ALL, GISSEL, PEGASE, PEGASE Sa13 in Table \ref{redshifts100}, 
\ref{redshifts20}, \ref{redshifts10} and \ref{redshifts6}) 
may appear surprising, as one would expect more
degeneracy when simulations include independent models of template SEDs.
However, we project the sub-mock catalogues onto the full realistic
catalog, which includes all template SEDs.
An interesting test is to measure, for instance,
redshifts from a GISSEL simulation using a PCA calculated from a
PEGASE-based mock catalog, and to examine whether significant errors
occur. Table \ref{redshiftzsys} shows the
systematic error thus introduced in the measurement of redshifts for
signal-to-noise ratios $S/N=100, 20, 10,$ and $6$; Fig{.} \ref{zsys.ps} shows
the measured (``observed'') redshift versus the true redshift (``theoretical'')
for $S/N=20$.  The standard
deviation $\sigma_\mathrm{Res.}$ in the residuals rises by an order of magnitude
for $S/N=100$ compared to the results in Table \ref{redshifts100}, and the average
deviations show a systematic negative offset. For lower signal-to-noise ratios
($S/N=20, 10, 6$), the degradation in the redshift measurement is less marked 
as the $S/N$ decreases, as expected; the systematic negative offset persists
and is largest for $S/N=6$. 
Table \ref{tablestargalzsys} shows the classification efficiency for this
simulation: a significant $6.6\%$ of galaxies are misidentified as
stars at a median $S/N=100$ and $3\%$ at a median $S/N=20$. The misidentification
occurs because the PEGASE templates do not fill the 10-D PCA space in the same way as GISSEL
templates do; the effect is smaller for lower $S/N$ because noise partly
dissipates the difference. 

This test therefore shows that classification and redshift
measurement are sensitive to the representativity of each
analyzed spectrum among the sample from which the PCA is performed:
a subsample of observed SEDs not represented in the catalog used for
the PCA might have systematically biased redshifts and classification.
This hints to the inaccuracies of the current spectral synthesis
models rather than to a real degeneracy problem for PCA, and urges the
use of the widest variety of galaxy/QSO/stellar SEDs when performing a
PCA. Instead of performing the PCA onto the observed sample, a better
method is to calculate the eigenvectors from a mock catalog with the widest
variety of galaxy, QSO, and stellar SEDs, together with a realistic
mix of the different object types and realistic redshift
distributions. This was attempted with the mock catalogues generated
here.  However, the systematic differences between the GISSEL and
PEGASE templates show that improved models of spectral libraries are
needed.

\begin{figure}
\begin{center}
\epsfysize=6cm
\centerline {\epsfbox[35 23 552 737]{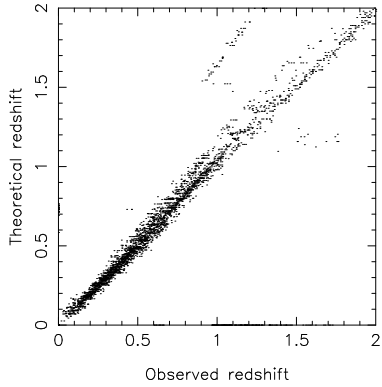}}
\caption{Measured redshift versus input redshift using
a $\chi^2$ reduction of the first 10 PCA eigencomponents of a GISSEL-based mock
catalog projected onto a PEGASE-based PCA. The median signal-to-noise ratio
is $S/N=20$. The corresponding standard deviations for $z\le1.7$ are
listed in Table \protect\ref{redshiftzsys}.}
\label{zsys.ps}
\end{center}
\end{figure}

\begin{table}
\begin{center}
\caption{Star/Galaxy/QSO classification efficiency for a GISSEL-based mock catalog
projected onto a PEGASE-based PCA (see text for details).}
\label{tablestargalzsys}
\begin{tabular}{lllll}
\hline
\hline
\multicolumn{5}{c}{Median $S/N = 100$}\\
\hline
Type& Input \# & \multicolumn{3}{c}{Output \#}\\
&&Star (\%)&Gal. (\%)&QSO (\%)\\
\hline
Stars &  3181 & 3181 (100) & 0 (0) & 0 (0)\\
Galaxies & 2993 & 198 (6.6) & 2795 (93.4) & 0 (0)\\
QSO & 100 & 0 (0) & 0 (0) & 100 (100)\\
\hline
\multicolumn{5}{c}{Median $S/N = 20$}\\
\hline
Type& Input \# & \multicolumn{3}{c}{Output \#}\\
&&Star (\%)&Gal. (\%)&QSO (\%)\\
\hline
Stars &  2843 & 2823 (99) & 20 (1) & 0 (0)\\
Galaxies & 2993 & 83 (2.8) & 2910 (97.2) & 0 (0)\\
QSO & 100 & 0 (0) & 0 (0) & 100 (100)\\
\hline
\multicolumn{5}{c}{Median $S/N = 10$}\\
\hline
Type& Input \# & \multicolumn{3}{c}{Output \#}\\
&&Star (\%)&Gal. (\%)&QSO (\%)\\
\hline
Stars &  2092 & 2092 (100) & 0 (0) &  (0)\\
Galaxies & 2993 & 29 (1) & 2964 (99) & 0 (0)\\
QSO & 984 & 0 (0) & 0 (0) & 984 (100)\\
\hline
\multicolumn{5}{c}{Median $S/N = 6$}\\
\hline
Type& Input \# & \multicolumn{3}{c}{Output \#}\\
&&Star (\%)&Gal. (\%)&QSO (\%)\\
\hline
Stars &  1956 & 1956 (100) & 0 (0.) & 0 (0)\\
Galaxies & 2095 & 2 (0.1) & 2093 (99.9) & 0 (0)\\
QSO & 706 & 0 (0) & 1 (0.1) & 704 (99.9)\\
\hline
\end{tabular}
\end{center}
\end{table}

\section{Discussion}\label{discussion}

\subsection{Degeneracy in PCA}\label{degeneracy}

\begin{figure}
\begin{center}
\epsfysize=5cm
\centerline {\epsfbox[0 0 550 500]{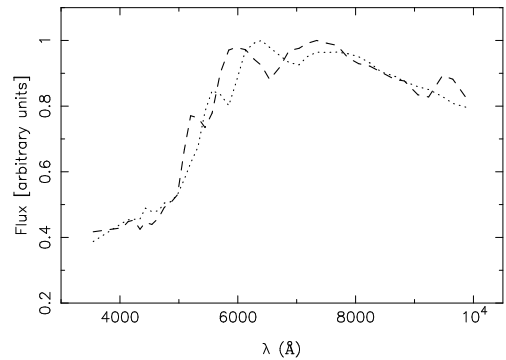}}
\caption{SEDs of 2 GISSEL SAB galaxies viewed through the LZT filter system
(see Fig{.} \protect\ref{transcurv}), with high $S/N$ and slightly different
IMF metallicity and redshifts: Salpeter IMF, metallicity [Fe/H]$=0.004$,
and $z=0.354$ for the SED shown as dotted lines; and Salpeter IMF, metallicity
[Fe/H]$=0.020$ and $z=0.274$ for the SED shown as dashed line. 
The 2 SEDs are clearly distinguishable.}
\label{SAB_0.354_0.274}
\end{center}
\end{figure}

\begin{figure}
\begin{center}
\epsfysize=5cm
\centerline {\epsfbox[0 0 550 500]{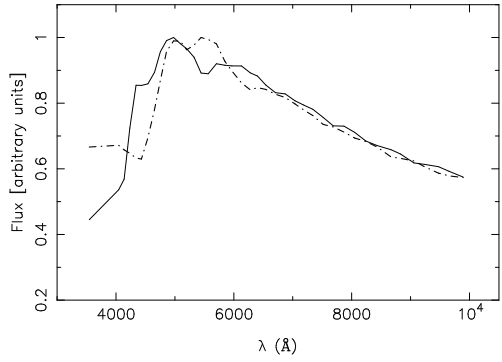}}
\caption{Same as in Fig{.} \ref{SAB_0.354_0.274} using the PEGASE library:
an old Sa galaxy at $z=0.204$ (solid line) and an Elliptical galaxy at 
$z=0.064$ (dot-dashed line).}
\label{Sa13_0.204_E13_0.064}
\end{center}
\end{figure}

We now examine the problem of degeneracies in the PCA, which
is particularly acute when too few eigencomponents are used.
Figures \ref{SAB_0.354_0.274}, \ref{Sa13_0.204_E13_0.064}, and
\ref{construct} illustrate the limit of reconstructing SEDs using
only the first 4 eigencomponents of the PCA. Figures
\ref{SAB_0.354_0.274} and \ref{Sa13_0.204_E13_0.064} show LZT SEDs of
galaxies which have different redshifts but similar continua. In
principles, the spectra should be clearly distinguishable using their
local features. But the 4-component PCA classifies them as ``similar'' spectra,
hence introducing a large error in the redshift measurement. Using
Eq{.} \ref{zmeas}, the PCA measures $z=0.354$ for the pair of SEDs in
Fig{.} \ref{SAB_0.354_0.274}, which have true redshifts $z=0.354$,
$z=0.274$, and $z=0.204$ for the pair in Fig{.}
\ref{Sa13_0.204_E13_0.064}, with true redshifts $z=0.204$, $z=0.064$.
Figure \ref{construct} shows the PCA reconstruction of the SEDs using
the first 4 eigencomponents and their corresponding eigenvectors.  The
reconstructed SEDs are nearly identical in each pair. The 4-component PCA washed out all local
features precluding any finer analysis. 
We also estimated the redshift measurement error
using only the 4-component PCA. The redshift accuracies at 
a median $S/N=10$ were 3 times larger than those obtained
with the 10-component PCA at the same $S/N$ ratio 
(see Table \ref{redshifts10}). 

The problem of degeneracies when trying to measure redshifts using
the PCA is linked to an intrinsic limit
of the technique. The PCA is powerful for filtering the noise in
observed spectra, for extracting both the common features 
and the major source of variance among an ensemble of SEDs.
In the case of galaxy SED's at zero redshift, the PCA is well 
suited for accurate classification, as the common features of the 
spectra are the underlying stellar populations and the characteristic
emission and absorption lines associated with them.
In the present case, the mix of redshifts which cover a large interval
($0\la z\la 2$ for galaxies) introduces a vast variety of continuum shapes among the SEDs
and the dominant eigencomponents of the resulting PCA emphasize the continuum
over local features, which are common to only a few objects
(those at the same redshift).

\begin{figure}
\begin{center}
\epsfysize=5cm
\centerline {\epsfbox[0 0 550 500]{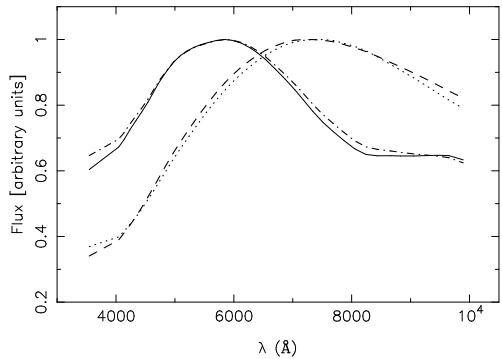}}
\caption{The reconstructed SEDs using the first 4 components of the
PCA are shown for the 4 galaxies of Fig{.} \protect\ref{SAB_0.354_0.274}
and \protect\ref{Sa13_0.204_E13_0.064} using the same line types. 
Differentiating between the 2 SAB galaxies, resp. between the Sa and 
the E galaxy, is intrinsically difficult.}
\label{construct}
\end{center}
\end{figure}

Most of the degeneracies are resolved when using 10 eigencomponents 
instead of 4, because 10 components allow one to reproduce the main
absorption bands and discontinuities through the entire range of
redshifts. Figures \ref{E13_1800_0.8_Sa13_12000_0.688.ps} and
\ref{E13_1800_0.8_Sa13_12000_0.688c.ps} illustrate the description
improvement of a 10-eigencomponents analysis. Figure 
\ref{E13_1800_0.8_Sa13_12000_0.688.ps}  shows
a PEGASE E template of 1.8 Gy at a redshift $z=0.8$ (solid
line) superposed on a PEGASE Sa spiral template of 12 Gy at a
 redshift $z=0.688$ (dot-dashed line).
Figure \ref{E13_1800_0.8_Sa13_12000_0.688c.ps} shows the PCA reconstruction
of the same templates using 10 eigencomponents. High signal-to-noise
ratio spectra would be distinguishable whereas low signal-to-noise
spectra might be misclassified.

\begin{figure}
\begin{center}
\epsfysize=5cm
\centerline {\epsfbox[0 0 550 500]{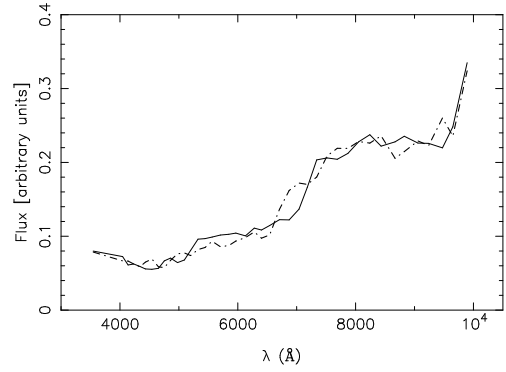}}
\caption{SEDs of 2 PEGASE galaxies viewed through the LZT filter system
(see Fig{.} \protect\ref{transcurv}), with intermediate $S/N$, and different ages and redshifts:
E with age 1.8 Gy, and $z=0.800$ (solid line),and Sa with age 12 Gy and $z=0.688$
(dot-dashed line); both have similar metallicities and IMF of Kroupa \protect\citet{fioc97}.
The 2 SEDs are clearly distinguishable as well as the PCA reconstructions in
Fig{.} \protect\ref{E13_1800_0.8_Sa13_12000_0.688c.ps}.}
\label{E13_1800_0.8_Sa13_12000_0.688.ps}
\end{center}
\end{figure}

\begin{figure}
\begin{center}
\epsfysize=5cm
\centerline {\epsfbox[0 0 550 500]{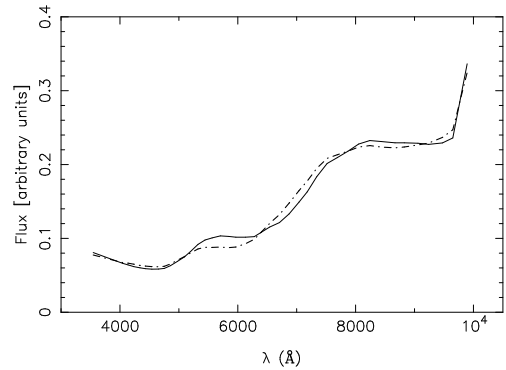}}
\caption{Reconstruction of the templates shown in
 Fig{.} \protect\ref{E13_1800_0.8_Sa13_12000_0.688.ps} using 10
eigencomponents of a PCA mock catalog.
High signal-to-noise ratio spectra projected onto such a PCA would
allow us to differentiate the 2 reconstructions, but low signal-to-noise
SEDs would be indistinguishable.}
\label{E13_1800_0.8_Sa13_12000_0.688c.ps}
\end{center}
\end{figure}

Such degeneracies have however a weak impact on the overall performances
of the PCA technique as indicated in Tables \ref{redshifts20} to \ref{redshifts6}.
Catastrophic degeneracies in redshift affect only 2\% of the galaxies, 
and 10\% of the QSOs at all signal-to-noise ratios considered; 
these fractions are calculated as the fractions of
objects in each class for which the difference between the input
and measured redshift is larger than the quadrature sum of the redshift 
residuals corresponding to the 2 redshift values (listed in Tables
\ref{redshifts20}--\ref{redshifts6}).
As the degeneracy sometimes only affect the redshift and not the spectral
type (see Fig{.} \ref{SAB_0.354_0.274}), the degeneracies in redshift yield 
degeneracies in spectral type for only a sub-fraction of the objects.

Note that if the catalog used to calculate the PCA is different
from the observed catalog for which the classification, type and
redshift measurement are planned to be performed, 
a pre-requisite is to introduce in the source
catalog for the PCA similar error functions as in the data. Projecting
an observed catalog with high signal-to-noise ratio, onto a PCA
calculated from low signal-to-noise spectra would merely result in the
loss of information present in the data. Whereas projecting an
observed catalog with low signal-to-noise onto a PCA derived from
high signal-to-noise data would result in catastrophic errors.

\subsection{Comparison with photometric redshifts}

\begin{figure}
\begin{center}
\epsfysize=5cm
\centerline {\epsfbox[0 0 550 500]{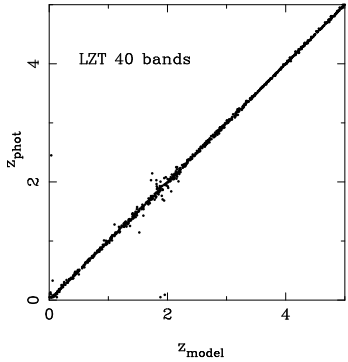}}
\caption{Comparison of the input redshifts $z_\mathrm{model}$ of a simulated catalog
and the photometric redshift estimation $z_\mathrm{phot}$ from HYPERZ code
\protect\citep{bolzonella00} when using the LZT 40 medium-band
filters. The redshift sequence is tight at all redshifts showing only
a small degeneracy near redshift 2.}
\label{plotphotom}
\end{center}
\end{figure}
\begin{figure}
\begin{center}
\epsfysize=5cm
\centerline {\epsfbox[0 100 550 600]{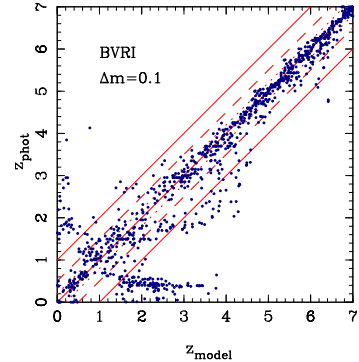}}
\caption{Same as in Fig{.} \protect\ref{plotphotBVRI} using $BVRI$ 
broad-band filters. This figure is reproduced from Fig{.} 2 of
\protect\citep{bolzonella00}; dotted lines correspond to $\Delta z=0.2$, dashed lines
to $\Delta z=0.5$, and thin solid lines to $\Delta z=1$, where
$Delta z=|z_\mathrm{model}-z_\mathrm{phot}|$. Conspicuous degeneracies appear along
the redshift sequence. }
\label{plotphotBVRI}
\end{center}
\end{figure}

The galaxy redshift accuracy $\sigma_\mathrm{Res.} \sim 0.01-0.05$
(for median $S/N\ge10$) obtained here with the PCA should
not be attributed to the PCA technique itself, but to the dense
sampling of the SEDs by the LZT medium-band filters (note that
$\sigma_\mathrm{Res.}\sim 1/R$ with the LZT filter resolution $R\simeq40$, see Fig{.} \ref{transcurv}).
We illustrate this point by using another method to measure the
redshifts of ``LZT-observed'' galaxies. We use the published code
HYPERZ \citep{bolzonella00} for measuring ``photometric redshifts'' by
standard $\chi^2$ fits to a spectral library. The model catalog of
SEDs was kindly provided by R. Pell\'o and is described in 
\citet{bolzonella00}: it contains the SEDs with a $S/N=10$ 
of 1000 galaxies generated using version GISSEL98 of the GISSEL spectral library
(Bruzual and Charlot, 1993; see Sect{.} 2.3.2). This sample was 
then ``observed'' through the LZT filters.  Figure \ref{plotphotom} shows a
comparison between the input redshifts $z_\mathrm{model}$ of the resulting
catalog with their photometric estimation $z_\mathrm{phot}$. In the redshift
range of interest for the LZT galaxies ($0\le z\le1.5$), the
r{.}m{.}s{.} dispersion in the redshifts errors
$\Delta z=|z_\mathrm{phot}-z_\mathrm{model}|$ (after rejection of the few
``catastrophic detections'', defined by $\Delta z\ge 1$, and occurring
only at $z \le0.5$), is $0.036$ in the interval $0\le z\le0.5$,
$0.018$ in the interval $0.5\le z\le1.0$, $0.033$ in the interval
$1.0\le z\le1.5$. Note that these dispersions are comparable with the
values obtained with the PCA in the full redshift range $0\le z\le1.5$ (see Table
\ref{redshifts10}).  

For comparison, we replicate in Fig{.} \ref{plotphotBVRI}, the $BVRI$
panel of Fig{.} 2 from \citet{bolzonella00},
showing the dispersion in the photometric redshifts calculated from
broad-band magnitudes with an uncertainty $\Delta m=0.1$ (which
approximately matches the $S/N=10$ in the LZT SEDs used in Fig{.}
\ref{plotphotom}): in the redshift range $0\le z \le1.5$, the r{.}m{.}s{.} 
uncertainty in $z_\mathrm{phot}$ is in the range $0.2-0.3$ 
\citep[see Table 2 of][]{bolzonella00}, 10 times larger than with
the LZT filters. The variance in the redshifts error is also
calculated after rejection of the ``catastrophic detections'' with
$\Delta z\ge 1$. These correspond to analogous degeneracies as those
described here in Sect{.} 6.1, and mainly occur in the redshift
intervals $0\le z\le0.4$ et $1.5\le z\le4.5$.  We emphasize that the PCA-measured redshift sequence
in Fig{.} \ref{plotphotom} is much cleaner than that obtained
with the broad-band filters in Fig{.} \ref{plotphotBVRI}. As both sets
of filters span the same wavelength range, this comparison 
provides a direct demonstration of the general advantage of using the
LZT medium-band filters over the traditional use of broad-band
filters, in proportion to the resolution of the filter system used,
and at the expense of observing time.  An intermediate approach in
which a wide-band system is complemented by medium-band filters was also
proposed by \citet{budavari01}, with a moderate
gain in redshift accuracy (the error is decreased by 34\% for
$z\le1.3$).

\subsection{Improvements}

The measurements of physical parameters using LZT SEDs might be improved
by complementing the PCA with a $\chi^2$ 
technique for measuring the redshifts. Based on spectral
libraries such as those used here, a $\chi^2$ analysis measures redshifts to
$\sigma_\mathrm{Res.}\la0.02$ at $z\la 1$ for $S/N=10$ \citep{hickson94a}. It is 
however very sensitive to the noise in the SEDs, as it makes use of all the 
information carried in them. This approach also provides a way to perform
the object classification (note that at redshift $z\la0.5$, comparison
of the object profiles with the point-spread-function could be used for
differentiating stars from galaxies, as discussed in Sect{.} \ref{extended}).
One could use the  $\chi^2$ technique to control the redshifts
calculated by the 10-component PCA, and thus identify the degenerate SEDs for
which better redshifts could be measured.
Alternatively, one could consider a 2-step analysis in which the $\chi^2$ technique
is first applied to the observed SEDs in order to measure their redshifts, and
in a second step, the PCA is applied
to the SEDs ``blue-shifted'' to $z=0$, with the goal to obtain a reliable
spectral classification for the identified objects, and to generate
``noiseless'' reconstructed SEDs. An iterative process using both 
techniques might be required and would contribute to improving the
object classification and redshift measurement.           

Because observing objects at widely different redshifts results in
rest-wavelength SEDs covering differing wavelength intervals. The
basic PCA technique described above and requiring that objects are
covered by a common wavelength interval, could not be used on the SEDs
``blue-shifted'' to $z=0$.  One could consider performing separate
PCAs on sub-samples of rest-wavelength SEDs selected by redshift
interval. Studying the evolution of the galaxy populations with
redshift would however require to compare the SEDs at different
redshift over a common wavelength interval.  A technique, recently
developed by Connolly and co-workers, uses the reconstructing ability
of the PCA to fill the gaps in data with varying wavelengths intervals
\citep{connolly99}. A complementary approach which creates the spectral
templates from multicolor surveys \citep{budavari00} could also be
applied.  Overall, the resulting PCA would allow both a reliable
spectral classification of all objects, and reconstruction of the
missing regions of the spectra under the null-evolution hypothesis
(which the PCA would implicitly do).  By comparison of how the
different spectral types are populated at different redshifts, such a
PCA would allow one to examine and quantify the redshift evolution of
the galaxy population.

\section{Conclusion}

This paper describes an application of Principal Component Analysis (PCA)
to a simulated multicolor survey using the 40 medium-band filters of the
Large Zenith Telescope. For that purpose, we generate realistic mock 
catalogues of $\sim3000$ stars, $\sim 30\,000$ galaxies, and $\sim 1\,000$ QSOs. 
For stars, we use templates 
from the library of Pickles (1998) and the phenomenological model 
of star counts of Bahcall \& Soneira (1986). For galaxies, we use spectral energy distributions (SED) 
from GISSEL (Charlot, 1993) and PEGASE (Fioc, 1997) and a luminosity function
derived from a review of the most recent $R$-band luminosity functions 
of the literature. We choose the CFRS-type evolution in the galaxy 
luminosity function, with luminosity evolution of only the late-type galaxies.
For QSOs, we use an extrapolation of the composite spectrum of Francis (1991), 
and the luminosity function of the 2dF QSO survey.

Using the realistic mock catalogues, we perform a PCA and extract the
first 10 eigencomponents. The 10-D space allows one to separate 
efficiently stars, galaxies, and QSOs even at low signal-to-noise 
ratios.  98\% of stars, 100\% of galaxies and 93\% of QSOs
are classified correctly at a median signal-to-noise ratio $S/N=6$. 
These values increase to 100\% of stars, 100\% of galaxies and 100\% of QSOs at 
a median $S/N=10$. For SEDs with a median $S/N=6$, the 10-component PCA 
also provides a measurement of redshifts accurate to $\sigma_\mathrm{Res.}\la0.05$ for galaxies 
with $z\la0.7$, and to $\sigma_\mathrm{Res.}\la0.2$ for QSOs with $z\ga2$. 
At a median $S/N=20$, $\sigma_\mathrm{Res.}\la0.02$ for galaxies 
with $z\la1$ and for QSOs with $z\ga2.5$ (for a given median $S/N$,
a 10 to 30 times lower $S/N$ is expected at the extreme wavelengths 
of the bluest/reddest objects).
This is not sufficient for small-scale 3-D clustering analyses,
but perfectly adequate for luminosity function studies, and for measuring
the evolution with redshift in the large-scale clustering using projected
moments.

This paper also underlines the main weakness of the PCA. It is
well-known that age, star-formation rate, redshift, and dust
extinction produce degenerate SEDs at a
resolution $R\simeq40$. Although the PCA is not able to resolve some
intrinsic degeneracies due to the medium-band observing technique, it
efficiently reduces the noise in the SEDs at the expense of additional
degeneracies. The solution to this problem may lie in the combination
of a PCA with a standard $\chi^2$ fitting procedure. Another crucial
issue in the use of the PCA for type/class/redshift measurement, is to
calculate the eigenvectors from a sample in which each type of object
is sufficiently well represented. The use of such a catalog, constructed
from a combination of a wide variety of well calibrated observed SEDs
together with precise evolutionary models, will guarantee the best
results for PCA analyses. These reference samples will also allow
detection of new types of object, as these will significantly deviate
from the sequences of known objects.

\begin{acknowledgements}

We wish to thank B. Rocca-Volmerange and M. Fioc for kindly providing
help with the public package PEGASE, S. Charlot for sending GISSEL95 templates
and insightful comments, R. Smith for information on the 2dF QSO, Roser Pell\'o
and Micol Bolzonella for providing a model catalog and for their 
efficient help using HYPERZ.
We are also grateful to the anonymous referee who helped us in improving
the content of this article.
This works has been made possible by a post-doctoral fellowships to RC
from the Fonds FCAR, Qu\'ebec.
\end{acknowledgements}

\bibliography{aamnem99,MS1340}
\bibliographystyle{aa}

\end{document}